\documentclass[lettersize,journal]{IEEEtran}

\usepackage{amsmath,amssymb,bm,mathtools}

\usepackage{graphicx}
\usepackage[caption=false,font=footnotesize,labelfont=rm,textfont=rm]{subfig}

\makeatletter
\renewcommand{\p@subfigure}{\thefigure}
\makeatother
\usepackage{stfloats}

\usepackage{array}

\usepackage{cite}
\usepackage{url}

\usepackage{mathrsfs}

\usepackage{xcolor}
\definecolor{revisionblue}{HTML}{0000FF}
\usepackage{color}
\usepackage{makecell}

\usepackage[ruled,linesnumbered]{algorithm2e}

\SetKwFor{ForE}{\textbf{for}}{\textbf{ do}}{\textbf{ end for}}
\SetKwFor{WhileE}{\textbf{while}}{\textbf{ do}}{\textbf{ end while}}
\SetKwIF{IfE}{ElseIfE}{ElseE}{\textbf{if}}{\textbf{ then}}{\textbf{ else if}}{\textbf{ else}}{\textbf{ end if}}

\SetAlFnt{\small}          
\SetAlCapFnt{\small}       
\SetAlCapNameFnt{\small}   
\SetAlgoNlRelativeSize{0}       
\SetInd{0.8em}{1.2em}           
\DontPrintSemicolon          

\usepackage{textcomp}
\usepackage{verbatim}
\usepackage{enumitem}
\RequirePackage{bibspacing}

\AtBeginDocument{
    \setlength{\jot}{0pt}
    \setlength{\abovedisplayskip}{0pt}
    \setlength{\abovedisplayshortskip}{0pt}
    \setlength{\belowdisplayskip}{0pt}
    \setlength{\belowdisplayshortskip}{0pt}
}
\definecolor{mygreen}{RGB}{252, 93, 3}   
\usepackage[colorlinks=true,
            linkcolor=mygreen,     
            anchorcolor=mygreen,
            citecolor=blue,      
            urlcolor=blue]{hyperref}

\definecolor{footnotepurple}{RGB}{0,204,102}

\makeatletter
\renewcommand{\@makefnmark}{%
  \hbox{\@textsuperscript{\normalfont\color{footnotepurple}\@thefnmark}}%
}
\makeatother

\usepackage{cleveref}
\crefname{algocf}{Algorithm}{Algorithms}
\Crefname{algocf}{Algorithm}{Algorithms}

\usepackage{xpatch}
\makeatletter
\ExplSyntaxOn
\cs_new:Npn \bibColoredItems #1#2
  { \clist_map_inline:nn {#2} { \cs_new:cpn {bib@colored@##1} {#1} } }
\ExplSyntaxOff
\newcommand\bib@setcolor[1]{%
  \ifcsname bib@colored@#1\endcsname
    \expanded{\noexpand\color{\csname bib@colored@#1\endcsname}}%
  \else
    \normalcolor
  \fi
}
\IfPackageLoadedTF{hyperref}{\@tempswatrue}{\@tempswafalse}
\if@tempswa
  \xpatchcmd\@bibitem {\H@item}{\bib@setcolor{#1}\H@item}{}{\PatchFailed}
  \xpatchcmd\@lbibitem{\H@item}{\bib@setcolor{#2}\H@item}{}{\PatchFailed}
\else
  \xpatchcmd\@bibitem {\item}  {\bib@setcolor{#1}\item}  {}{\PatchFailed}
  \xpatchcmd\@lbibitem{\item}  {\bib@setcolor{#2}\item}  {}{\PatchFailed}
\fi
\makeatother

\definecolor{revisionblue}{HTML}{0000FF}

\begin{document}

\title{Traffic-Adaptive Per-Hop Multipath Routing in Multi-Hop UAV Networks}

\author{Zhenyu Zhao,~\IEEEmembership{Graduate Student Member, IEEE}, Tiankui Zhang,~\IEEEmembership{Senior Member, IEEE}, Xiaoxia Xu,~\IEEEmembership{Member, IEEE}, Yuanpeng Zheng, Junjie Li, Wenjuan Xing
\thanks{Zhenyu Zhao and Tiankui Zhang are with the School of Information and Communication Engineering, Beijing University of Posts and Telecommunications, Beijing 100876, China (e-mail: zhaozhenyu@bupt.edu.cn; zhangtiankui@bupt.edu.cn).}

\thanks{Yuanpeng Zheng is with the China Mobile Zijin (Jiangsu) Innovation Research Institute, Jiangsu 210033, China (e-mail: zhengyp@js.chinamobile.com).}

\thanks{Xiaoxia Xu is with the School of Electronic Engineering and Computer Science, Queen Mary University of London, London E1 4NS, U.K. (e-mail: x.xiaoxia@qmul.ac.uk).}

\thanks{Junjie Li is with China Telecom Beijing Research Institute, Beijing 102200, China (e-mail: lijj28@chinatelecom.cn).}

\thanks{Wenjuan Xing  is with the School of Microelectronics and Communication Engineering, Chongqing University, Chongqing 401331, China. (e-mail: xingwj@chinatelecom.cn).}

}



\maketitle

\begin{abstract}

In uncrewed aerial vehicle (UAV)-relayed mobile edge computing (MEC) networks, computation tasks generate traffic with diverse latency requirements and data sizes. Routing decisions therefore need to adapt to both traffic characteristics and changing network conditions. Compared with single-path routing, multipath routing is better suited to such heterogeneous traffic because it provides multiple forwarding options and enables flexible traffic splitting. However, conventional multipath routing usually splits traffic over predefined end-to-end paths, making it difficult to respond quickly to link fluctuations and topology changes in UAV networks. To address this issue, we propose a traffic-adaptive per-hop multipath routing method for multi-hop UAV networks, in which each UAV dynamically distributes traffic among multiple candidate next hops. We formulate the routing problem to improve the on-time packet delivery ratio while reducing the packet loss ratio, and model it as a decentralized partially observable Markov decision process (Dec-POMDP). To solve this problem, we develop a multi-agent reinforcement learning (MARL) algorithm, termed Multi-Agent Proximal Policy Optimization with Dirichlet Modeling (MAPPO-DM). MAPPO-DM follows the centralized-training-and-decentralized-execution framework and models continuous traffic-splitting actions using a Dirichlet distribution. Simulation results show that MAPPO-DM outperforms the baseline methods and maintains robust performance under various network conditions.

\end{abstract}

\begin{IEEEkeywords}
Uncrewed aerial vehicle (UAV)-relayed mobile edge computing (MEC), per-hop multipath routing, multi-agent reinforcement learning (MARL).
\end{IEEEkeywords}

\section{Introduction}
\IEEEPARstart{U}{ncrewed} aerial vehicles (UAVs) possess advantages such as rapid deployment, wide coverage, and flexible networking~\cite{UAV-intro-1, UAV-intro-2}, making them a promising solution for enhancing communication and edge computing capabilities. In environments where ground infrastructure is limited, such as post-disaster areas or remote regions, UAVs can assist in establishing temporary communication networks~\cite{UAV-BS-1,UAV-BS-2} or mobile edge computing (MEC) networks~\cite{UAV-MEC-1,UAV-MEC-2}, providing essential communication and computing services. \par

In MEC networks, UAVs can take on different roles depending on application requirements. They may serve as aerial computing nodes to process computation tasks. Alternatively, they can act as relay nodes and forward data to ground base stations (GBSs) or other terminals through multi-hop transmission. When used as relay nodes, UAVs require flexible and efficient routing strategies to ensure the reliable delivery of computation-task traffic~\cite{UAV-routing-des-BS-1, UAV-routing-des-BS-2}. However, reliable data transmission remains challenging due to heterogeneous task traffic, frequent topology changes, and limited network resources. To address these challenges, various routing schemes have been developed for multi-hop UAV networks~\cite{UAV-OLSR-2, UAV-AODV-routing-reactive,TA-AOMDV, I-AOMDV, Ant-Colony-routing-hybrid,geo-routing,UAV-OLSR-1,q-learning-routing-1,q-learning-routing-2,q-learning-routing-3, swarm-learning-routing-1,swarm-learning-routing-2,trusted-learning-routing-1,trusted-learning-routing-2,uav-satellite-learning-routing,multipath-routing-satellite}. \par

\subsection{Related Work}

Existing routing studies for multi-hop UAV networks can be broadly divided into protocol-based and learning-based schemes. Protocol-based schemes establish and maintain transmission paths using predefined routing rules. They are widely used because of their clear decision logic and low implementation complexity~\cite{UAV-OLSR-2,UAV-AODV-routing-reactive,TA-AOMDV,I-AOMDV,Ant-Colony-routing-hybrid,geo-routing,UAV-OLSR-1}. In contrast, learning-based schemes use data-driven methods, especially reinforcement learning, to adapt routing decisions to changing network conditions. This makes them more flexible in complex and highly dynamic environments~\cite{q-learning-routing-1,q-learning-routing-2,q-learning-routing-3,swarm-learning-routing-1,swarm-learning-routing-2,trusted-learning-routing-1,trusted-learning-routing-2,uav-satellite-learning-routing,multipath-routing-satellite}.

\subsubsection{Protocol-based routing approaches}

According to~\cite{UAV-routing-survey}, protocol-based routing can be broadly divided into topology-based routing, geographic routing, and topology--geographic hybrid routing. Topology-based routing relies mainly on network topology information and establishes routes using proactive, reactive, or proactive--reactive hybrid mechanisms~\cite{UAV-OLSR-2,UAV-AODV-routing-reactive,TA-AOMDV,I-AOMDV,Ant-Colony-routing-hybrid}. Geographic routing makes forwarding decisions based mainly on node location information~\cite{geo-routing}. Topology--geographic hybrid routing combines topology information with geographic information to improve routing decisions~\cite{UAV-OLSR-1}.\par

In topology-based routing, proactive protocols periodically exchange control messages to maintain network topology information and select routes accordingly. Optimized Link State Routing (OLSR) is a representative proactive protocol. In~\cite{UAV-OLSR-2}, a neural network is incorporated into OLSR to consider link stability, bandwidth, and energy consumption during route selection, thereby improving routing stability. In contrast, reactive protocols discover routes only when data transmission is required. Ad hoc On-Demand Distance Vector (AODV) and Ad hoc On-Demand Multipath Distance Vector (AOMDV) are representative reactive protocols. AODV typically establishes a single end-to-end path between a source and a destination, whereas AOMDV maintains multiple loop-free candidate paths to improve reliability when links fail. In~\cite{UAV-AODV-routing-reactive}, the Technique for Order Preference by Similarity to Ideal Solution (TOPSIS) is integrated into AODV. The resulting route selection jointly considers hop count, link stability, and residual energy. In~\cite{TA-AOMDV}, residual energy, available bandwidth, queue length, and link stability are incorporated into AOMDV to construct multiple quality-of-service (QoS)-aware end-to-end paths. Among these paths, the most stable one is selected as the primary route. Similarly,~\cite{I-AOMDV} defines a service-aware QoS metric based on hop count, average nodal connectivity, and minimum available bandwidth. Several highly ranked end-to-end paths are then retained as primary and backup routes. Topology-based routing can also combine proactive and reactive mechanisms. Such proactive--reactive hybrid schemes maintain topology information within a local region for fast forwarding, while using on-demand route discovery for communication beyond that region. In~\cite{Ant-Colony-routing-hybrid}, a distributed hybrid ant-colony routing protocol combines proactive pheromone maintenance with reactive ant-based path discovery. \par

Geographic routing uses node locations to make hop-by-hop forwarding decisions. A packet is typically forwarded to a neighboring node that is closer to the destination or provides a higher forwarding utility. In~\cite{geo-routing}, a three-dimensional geographic routing protocol based on the effective transmission range is proposed. The next-hop node is selected according to link utility, while a three-dimensional perimeter recovery mechanism is used when relay candidates are unavailable.\par

Topology--geographic hybrid routing combines topology-based route maintenance with geographic information. In~\cite{UAV-OLSR-1}, node position, velocity, and moving direction are incorporated into the multipoint relay selection process of OLSR to improve routing stability.

\subsubsection{Learning-based routing approaches}

Learning-based routing schemes usually employ machine learning methods, such as reinforcement learning, to optimize routing decisions. Existing studies have explored several directions, including hop-by-hop routing, routing coordination in large-scale UAV swarms, secure routing, and routing in space--air integrated networks~\cite{q-learning-routing-1,q-learning-routing-2,q-learning-routing-3,swarm-learning-routing-1,swarm-learning-routing-2,trusted-learning-routing-1,trusted-learning-routing-2,uav-satellite-learning-routing,multipath-routing-satellite}.\par

For hop-by-hop routing, learning-based methods use information from neighboring UAVs, such as residual energy and link quality, to select the next-hop node. The studies in~\cite{q-learning-routing-1,q-learning-routing-2,q-learning-routing-3} develop Q-learning-based routing methods with reward functions that account for such neighbor information. These methods select suitable next hops to reduce end-to-end delay and balance energy consumption.\par

For routing in UAV swarm networks,~\cite{swarm-learning-routing-1} combines multi-agent reinforcement learning (MARL) with an adaptive communication mechanism to reduce flooding overhead and improve routing coordination. Similarly,~\cite{swarm-learning-routing-2} develops an intelligent cluster-based routing scheme to improve routing stability and load balancing through multi-factor weight optimization and adaptive cluster-head selection.\par

For scenarios with high security requirements,~\cite{trusted-learning-routing-1} develops a trust-aware routing framework that integrates blockchain with MARL. In~\cite{trusted-learning-routing-2}, a reinforcement-learning-based routing recovery scheme is proposed to restore network connectivity under targeted routing attacks \par

Learning-based routing has also been studied in space--air integrated networks to support reliable cross-domain transmission. For example,~\cite{uav-satellite-learning-routing} develops a two-layer deep reinforcement learning framework for UAV--satellite integrated Internet of Things networks. The framework combines hierarchical decision-making with a QoS-aware reward function to support reliable long-distance transmission with low latency. Related learning-based routing techniques have also been investigated in satellite networks. In~\cite{multipath-routing-satellite}, a graph neural network (GNN)-based multipath traffic-splitting method is proposed for low-Earth-orbit satellite networks. A centralized controller first generates multiple end-to-end candidate paths based on a global topology snapshot and then dynamically adjusts the traffic-splitting ratios according to current network conditions.

\subsection {Motivations and Challenges}

\subsubsection{Motivations}
Existing studies on multi-hop UAV routing have made significant progress in optimizing general performance metrics, such as energy efficiency and average latency. However, routing for heterogeneous computation-task traffic under high network loads remains largely unexplored. \par

Specifically, routing decisions should adapt to heterogeneous traffic with different latency requirements and data volumes. For latency-sensitive flows, routing should prioritize timely delivery to reduce end-to-end delay. For data-intensive flows, traffic should be balanced across the network to avoid packet loss caused by link congestion and queue buildup. Compared with single-path routing, multipath routing provides more forwarding options and allows traffic to be distributed across multiple paths according to service requirements and network conditions. It is therefore better suited to heterogeneous traffic, especially under heavy network loads or frequent link fluctuations.\par

However, existing multipath routing schemes usually construct multiple end-to-end paths based on network topology and then perform path selection, backup switching, or traffic splitting among these paths~\cite{I-AOMDV,TA-AOMDV,multipath-routing-satellite}. Such path-level designs are less responsive in highly dynamic multi-hop UAV networks, where UAV mobility can frequently change link conditions and even invalidate existing paths. Per-hop multipath routing provides a more adaptive alternative. Instead of relying on precomputed end-to-end paths, each UAV dynamically distributes traffic among multiple next-hop neighbors based on current link and traffic conditions. This enables faster adaptation to network changes and improves load balancing and transmission robustness. Nevertheless, per-hop multipath routing in multi-hop UAV networks has received limited attention.

\subsubsection{Challenges}

Designing traffic-adaptive per-hop multipath routing for multi-hop UAV networks remains challenging. The main challenges are as follows:

\begin{itemize}

\item \textit{Coupling between next-hop selection and traffic splitting:} Each UAV must select suitable next-hop neighbors based on current network conditions and determine how much traffic to assign to each selected neighbor according to traffic characteristics. These two decisions are closely coupled, leading to a large decision space and making efficient routing policies difficult to obtain.

\item \textit{Limited observability in per-hop routing:} Each UAV makes forwarding decisions using only local observations. However, a locally favorable decision may not lead to good end-to-end performance, especially for latency-sensitive traffic. Therefore, the routing policy should use local information effectively while accounting for its impact on end-to-end delay and overall network performance.

\end{itemize}

\subsection{Contributions}

This paper proposes a traffic-adaptive per-hop multipath routing method for heterogeneous computation-task traffic in multi-hop UAV networks. Each UAV uses local network and traffic information to dynamically split traffic among multiple next hops. We formulate the routing problem to jointly improve the on-time packet delivery ratio and reduce the packet loss ratio. The problem is modeled as a Decentralized Partially Observable Markov Decision Process (Dec-POMDP) and solved using a multi-agent reinforcement learning algorithm termed Multi-Agent Proximal Policy Optimization with Dirichlet Modeling (MAPPO-DM).\par

To the best of our knowledge, this is the first study to investigate traffic-adaptive per-hop multipath routing in multi-hop UAV networks. The main contributions are summarized as follows:

\begin{itemize}

\item We propose a traffic-adaptive per-hop multipath routing framework for heterogeneous computation-task traffic in multi-hop UAV networks. A priority-based sub-queue mechanism organizes traffic according to deadline urgency. Each UAV selects packets from the highest-priority non-empty sub-queue and dynamically splits them among multiple next hops. The routing problem is formulated to jointly improve the on-time packet delivery ratio and reduce the packet loss ratio.

\item We model the per-hop multipath routing problem as a Dec-POMDP and develop a MARL algorithm termed MAPPO-DM. The decentralized Actor uses a Transformer to model interactions among candidate next hops and a Gated Recurrent Unit (GRU) to capture temporal changes in local network conditions. A Dirichlet distribution is used to model continuous traffic-splitting actions, with its parameters adaptively adjusted to improve policy exploration. The centralized Critic uses graph attention to capture the network topology and node--link interactions from the global graph state, improving value estimation.

\item We evaluate MAPPO-DM under different traffic loads, network sizes, and candidate next-hop constraints. The comparison includes four representative baselines: two heuristic per-hop multipath methods, one per-hop multipath method guided by an existing end-to-end multipath routing algorithm~\cite{I-AOMDV}, and one heuristic per-hop single-path method. The results show that multipath routing generally outperforms single-path routing. Among the multipath methods, MAPPO-DM achieves the best overall performance and demonstrates strong adaptability and robustness.

\end{itemize}

The remainder of this paper is organized as follows. Section II presents the system model and problem formulation. Section III describes the proposed solution. Section IV presents the numerical results. Section V concludes the paper, and Section VI provides the acknowledgment.\par

\section{ System Model And Problem Formulation}
\subsection{System model}

\begin{figure}
    \centering
    \includegraphics[width=0.8\linewidth]{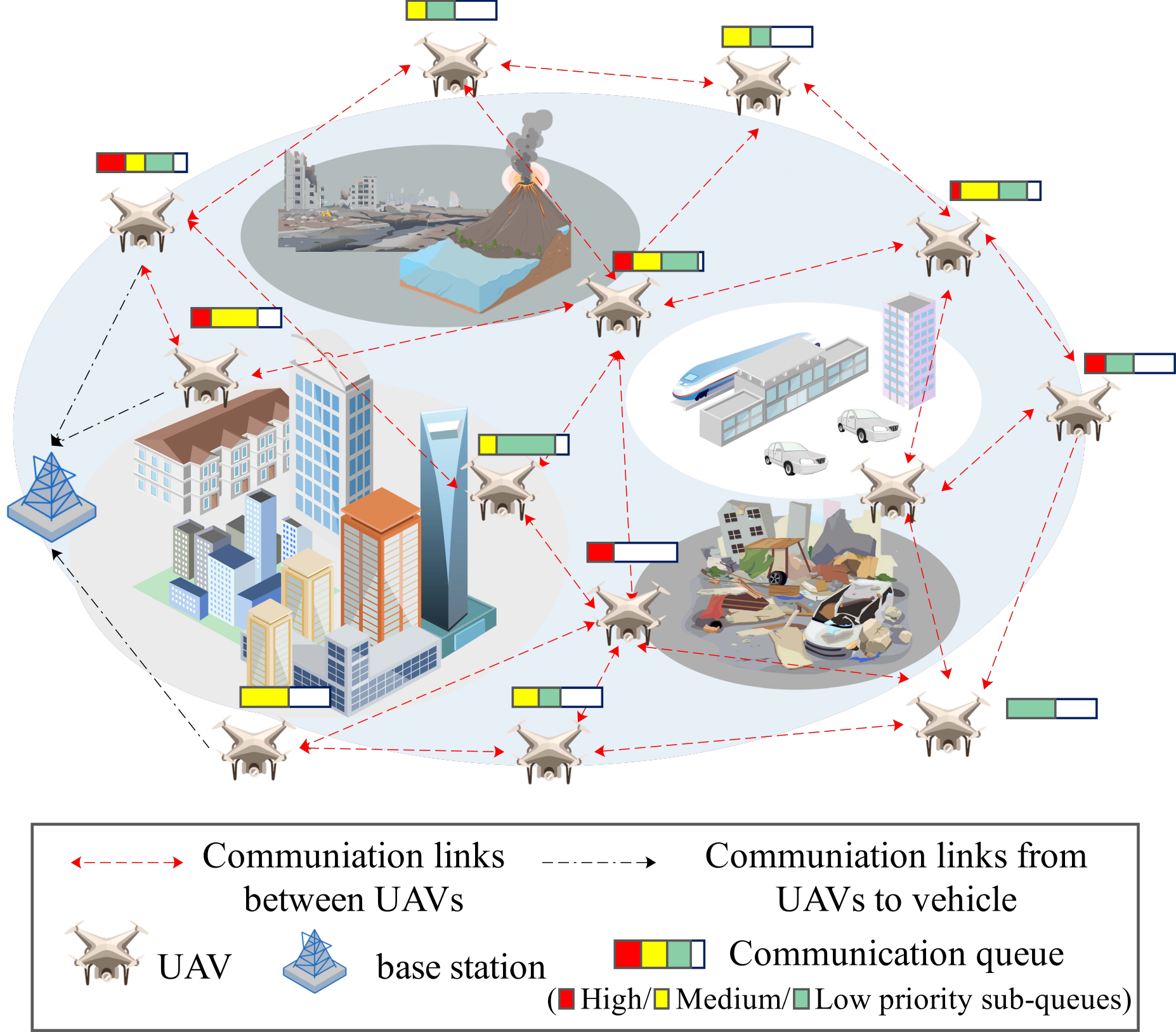}
    \caption{Multi-UAV and GBS Collaborative System.}
    \label{system-model}
\end{figure}

As shown in Fig.~\ref{system-model}, we consider a network where multiple UAVs cooperate with a GBS. The UAVs perform sensing or inspection tasks, generating computation-task traffic that needs to be transmitted to the GBS for processing and analysis.\par

The GBS is denoted by $k$, and the UAV set is ${\cal M}=\{1,\ldots,M\}$. Based on their locations, task-generation probabilities, and forwarding roles, the UAVs are divided into four disjoint subsets,
\begin{equation}
{\cal M} = {\cal M}_{\rm hot} \cup {\cal M}_{\rm gat} \cup {\cal M}_{\rm rel} \cup {\cal M}_{\rm reg},
\end{equation}
where ${\cal M}_{\rm hot}$, ${\cal M}_{\rm gat}$, ${\cal M}_{\rm rel}$, and ${\cal M}_{\rm reg}$ denote hotspot, gateway, relay, and regular UAVs, respectively. Their task-generation probabilities are denoted by $p_{\rm hot}^{\rm gen}$, $p_{\rm gat}^{\rm gen}$, $p_{\rm rel}^{\rm gen}$, and $p_{\rm reg}^{\rm gen}$, respectively. Hotspot UAVs operate in task-intensive areas and have the highest task-generation probability. Gateway UAVs are clustered near the GBS and mainly aggregate traffic for forwarding to the GBS, while generating little traffic themselves. Relay and regular UAVs are distributed throughout the operating area. Relay UAVs have a relatively low task-generation probability and mainly support multi-hop forwarding, whereas regular UAVs generate tasks more frequently. Accordingly, these probabilities satisfy $p_{\rm hot}^{\rm gen} > p_{\rm reg}^{\rm gen} > p_{\rm rel}^{\rm gen} > p_{\rm gat}^{\rm gen}.$\par

Let $T$ denote the total UAV flight duration. During this period, each UAV follows a predefined trajectory. The flight duration is divided into $T_{\rm max}$ equal time slots, each with length $\Delta t=T/T_{\rm max}$. The set of time slots is denoted by ${\cal T}=\{1,\ldots,T_{\rm max}\}$. We assume that $\Delta t$ is sufficiently small so that the UAV positions remain approximately unchanged within each slot, i.e.,
\begin{equation}\label{time-slot-constraint}
\Delta t \le \Delta t^{\max}.
\end{equation} The position of UAV $m$ in time slot $t$ is denoted by ${\bm l}_m(t)=[x_m^{\rm uav}(t),y_m^{\rm uav}(t),z_m^{\rm uav}(t)]$, while the GBS is fixed at ${\bm l}_k=[x_k^{\rm gbs},y_k^{\rm gbs},0]$.\par

Under the above time-slot model, each UAV may generate computation-task traffic in each time slot according to the probability associated with its role. Let $U$ denote the total number of traffic flows generated during $T$, with ${\cal U}=\{1,\ldots,U\}$. For each flow $u\in{\cal U}$, $t_u^{\rm end}$ and $I_u$ denote its deadline and data volume, respectively. Each flow is divided into fixed-size packets of size $I_{\rm pkt}$. Let $W_u$ denote the number of packets in flow $u$, with ${\cal W}_u=\{1,\ldots,W_u\}$. Each packet inherits the deadline of its corresponding flow, i.e., $t_w^{\rm end}=t_u^{\rm end}$ for $w\in{\cal W}_u$. The set of all packets generated during $T$ is denoted by ${\cal W}=\bigcup_{u\in{\cal U}}{\cal W}_u$.\par

\emph{1) Communication Model:} Since UAV-to-UAV links are generally dominated by line-of-sight (LoS) propagation, we adopt an LoS channel model for inter-UAV communications. In time slot $t$, the distance between UAVs $m$ and $m'$ is $d_{m,m'}(t)=\left\|{\bm l}_m(t)-{\bm l}_{m'}(t)\right\|_2.$ Let $\beta_0$ denote the channel gain at a reference distance of 1 m. The channel gain between UAVs $m$ and $m'$ is then given by $h_{m,m'}(t)=\frac{\beta_0}{d_{m,m'}^2(t)}.$\par

The system has $B$ orthogonal sub-bands, each with bandwidth $b_{\rm wid}$, and ${\cal B}=\{1,\ldots,B\}$ denotes the sub-band set. Each transmission link occupies one sub-band, while the same sub-band can be reused by different UAVs. Let $N_0$ denote the noise power spectral density. Each UAV maintains at most $N$ candidate next-hop UAV links and one direct link to the GBS. For simplicity, the transmit-power budget is equally divided among these $N+1$ possible outgoing links. Thus, the per-link transmit power of UAV $m$ is $p_m^{\rm tx}=\frac{p_m^{\rm tx,\max}}{N+1},$ where $p_m^{\rm tx,\max}$ is the maximum transmit power of UAV $m$.\par

When UAV $m$ transmits to UAV $m'$ on a given sub-band, let $\phi_{m'}(t)$ denote the set of UAVs transmitting on the same sub-band and causing interference at UAV $m'$. The resulting interference is $\ell_{m,m'}(t)
=
\sum_{i\in\phi_{m'}(t),\,i\neq m}
p_i^{\rm tx}h_{i,m'}(t).$ The corresponding signal-to-interference-plus-noise ratio (SINR) is $\vartheta_{m,m'}(t)
=
\frac{p_m^{\rm tx}h_{m,m'}(t)}
{b_{\rm wid}N_0+\ell_{m,m'}(t)}.$ \par

To ensure reliable and loop-free forwarding, UAV $m$ only considers UAV neighbors that satisfy both the SINR requirement and the geographic progress condition. Its reachable neighbor set is defined as $\psi_m(t)=\{m'\in{\cal M}\setminus\{m\}|\vartheta_{m,m'}(t)\geq\vartheta^{\min},d_{m',k}(t)<d_{m,k}(t)\},$ where $\vartheta^{\min}$ is the minimum SINR required for reliable transmission. The distance condition ensures that each hop moves packets closer to the GBS and therefore prevents routing loops.\par

For each reachable neighbor $m'\in\psi_m(t)$, we define the geographic forwarding score as $d^{\rm score}_{m,m'}(t)
=
\frac{d_{m,k}(t)-d_{m',k}(t)}
{d_{m,m'}(t)}.$ A larger score indicates greater progress toward the GBS relative to the forwarding distance. The UAVs in $\psi_m(t)$ are ranked in descending order of $d^{\rm score}_{m,m'}(t)$. If two UAVs have the same score, the one with the smaller identifier is ranked first. UAV $m$ retains at most the first $N$ UAVs as its candidate next-hop set, denoted by $\psi_m^{\rm com}(t)$.\par

For UAV $m$ and a candidate next hop $m'\in\psi_m^{\rm com}(t)$, the transmission rate is
\begin{equation}
R_{m,m'}(t)
=
b_{\rm wid}\log_2\left(1+\vartheta_{m,m'}(t)\right),
\quad
m'\in\psi_m^{\rm com}(t).
\label{comm-UAVs}
\end{equation}
Accordingly, the maximum number of packets that can be transmitted over this link in one time slot is
\begin{equation}
g_{m,m'}(t)
=
\left\lfloor
\frac{R_{m,m'}(t)\Delta t}{I_{\rm pkt}}
\right\rfloor,
\end{equation}
where $\lfloor\cdot\rfloor$ denotes the floor function.\par

Unlike UAV-to-UAV links, UAV-to-GBS links may experience both LoS and non-line-of-sight (NLoS) propagation because of ground obstacles. In time slot $t$, the distance and elevation angle between UAV $m$ and GBS $k$ are given by $d_{m,k}(t)=\left\|{\bm l}_m(t)-{\bm l}_k\right\|_2$ and $\theta_{m,k}(t)=\frac{180^\circ}{\pi}\arcsin\left(\frac{z_m^{\rm uav}(t)}{d_{m,k}(t)}\right)$, respectively. The LoS probability is modeled as $P_{m,k}^{\rm LoS}(t)
=
\frac{1}
{1+D_1\exp\left[-D_2\left(\theta_{m,k}(t)-D_1\right)\right]},$ where $D_1$ and $D_2$ are environment-dependent parameters. The NLoS probability is $P_{m,k}^{\rm NLoS}(t)=1-P_{m,k}^{\rm LoS}(t)$.\par

Following~\cite{LosNLos-channel-model}, the path loss under channel condition $\xi$ is $L_{m,k}^{\xi}(t)
= \eta_{\xi} \left(
\frac{4\pi f_c d_{m,k}(t)}{D_3}
\right)^{\hbar},\xi\in\{\rm LoS,NLoS\},$ where $\hbar$ is the path-loss exponent, $\eta_{\xi}$ is the excessive path-loss factor, $f_c$ is the carrier frequency, and $D_3$ is the speed of light. Based on the average path loss over the LoS and NLoS conditions, the effective channel gain is $h_{m,k}(t)
= \frac{1} {P_{m,k}^{\rm LoS}(t)L_{m,k}^{\rm LoS}(t)+P_{m,k}^{\rm NLoS}(t)L_{m,k}^{\rm NLoS}(t)}.$ \par

When UAV $m$ transmits to the GBS, let $\phi_k(t)$ denote the set of UAVs transmitting on the same sub-band and causing interference at the GBS. The interference is $\ell_{m,k}(t)
= \sum_{i\in\phi_k(t),\,i\neq m}
p_i^{\rm tx}h_{i,k}(t),$ and the corresponding SINR is $\vartheta_{m,k}(t)
=
\frac{p_m^{\rm tx}h_{m,k}(t)}
{b_{\rm wid}N_0+\ell_{m,k}(t)}.$ A direct UAV-to-GBS link is available only when $\vartheta_{m,k}(t)\geq\vartheta^{\min}$. For an available link, the transmission rate is
\begin{equation}
R_{m,k}(t)
=
b_{\rm wid}\log_2\left(1+\vartheta_{m,k}(t)\right).
\label{comm-uav-GBS}
\end{equation}
The maximum number of packets that UAV $m$ can transmit to the GBS within one time slot is then
\begin{equation}
g_{m,k}(t)
=
\left\lfloor
\frac{R_{m,k}(t)\Delta t}{I_{\rm pkt}}
\right\rfloor.
\end{equation}
\par

\emph{2) Queueing Model:} Each UAV is equipped with a finite-capacity communication buffer $Q_m$, whose maximum storage capacity is denoted by $q_m^{\max}$ (in packets). \par

To prioritize latency-sensitive traffic, the communication queue of each UAV is divided into three sub-queues corresponding to high, medium, and low priority levels, denoted as $Q_m(t) = \{ Q_m^1(t), Q_m^2(t), Q_m^3(t)\},$ where $Q_m^1(t)$, $Q_m^2(t)$, and $Q_m^3(t)$ represent the high-, medium-, and low-priority sub-queues, respectively. Let $q_m^1(t)$, $q_m^2(t)$, and $q_m^3(t)$ denote the numbers of packets stored in the corresponding sub-queues, respectively. Moreover, let $q_m^{\rm free}(t)$ denote the remaining buffer capacity of UAV $m$ at time slot $t$. These variables satisfy
\begin{equation}
    q_m^1(t) + q_m^2(t) + q_m^3(t) + q_m^{\rm free}(t) = q_m^{\max},  \forall m,t.
    \label{queue-capa-constr}
\end{equation}

At each time slot $t$, UAV $m$ selects the highest-priority non-empty sub-queue for packet forwarding, which is denoted by $Q_m^{\rm sel}(t)$. The corresponding priority level and queue length are represented by $\nu_m^{\rm sel}(t)\in\{1,2,3\}$ and $q_m^{\rm sel}(t)$, respectively.\par

Incoming or locally generated traffic is placed into different sub-queues according to their deadlines. Specifically, at time slot $t$, the priority of traffic $u \in \mathcal{U}$ is determined by its urgency, denoted as $u_{\rm prio}(t)$, and is given by
\begin{equation}
u_{\rm prio}(t) = 
\begin{cases} 
1, & \text{if } t_u^{\rm end} - t \Delta t \leq 4.5 \, \text{s}, \\ 
2, & \text{if } 4.5 \, \text{s} < t_u^{\rm end} - t \Delta t \leq 9 \, \text{s}, \\ 
3, & \text{if } t_u^{\rm end} - t \Delta t > 9 \, \text{s}.
\end{cases}
\label{traffic_prio}
\end{equation}

\emph{3) Forwarding Model:} In each time slot $t$, UAV $m$ forwards packets from its selected queue $Q_m^{\rm sel}(t)$. If a direct link to GBS $k$ is available, UAV $m$ transmits up to the UAV-to-GBS link capacity. Thus, the number of packets delivered to the GBS is $\min\{q_m^{\rm sel}(t),g_{m,k}(t)\}$. Packets that cannot be transmitted within the current time slot remain in the local queue for subsequent transmission.\par

If UAV $m$ cannot directly communicate with the GBS, it splits the packets in $Q_m^{\rm sel}(t)$ among multiple candidate next-hop UAVs and may retain some packets locally. The traffic-splitting action is defined as ${\bm a}_m(t)=\big[a_{m,m'}(t)\big]_{m'\in\{0\}\cup\psi_m^{\rm com}(t)}$, where $a_{m,0}(t)$ is the fraction retained locally and $a_{m,m'}(t)$ is the fraction assigned to candidate next-hop UAV $m'\in\psi_m^{\rm com}(t)$. The splitting ratios satisfy
\begin{align}
&0\leq a_{m,m'}(t)\leq1,
\quad \forall m'\in\{0\}\cup\psi_m^{\rm com}(t),\,t\in{\cal T},
\label{forward-0-1}\\
&\sum\nolimits_{m'\in\{0\}\cup\psi_m^{\rm com}(t)}
a_{m,m'}(t)=1,
\quad \forall t\in{\cal T}.
\label{forward-sum}
\end{align}\par

For each candidate next-hop UAV $m'\in\psi_m^{\rm com}(t)$, let ${\cal W}_{m,m'}^{\rm ori}(t)$ denote the set of packets assigned by UAV $m$ to UAV $m'$ in time slot $t$. Its cardinality is $|{\cal W}_{m,m'}^{\rm ori}(t)|=\operatorname{round}\big(a_{m,m'}(t)q_m^{\rm sel}(t)\big)$, where $\operatorname{round}(\cdot)$ denotes nearest-integer rounding.\par

Let ${\cal W}_{m,m'}^{\rm trans}(t)\subseteq{\cal W}_{m,m'}^{\rm ori}(t)$ denote the set of packets successfully transmitted from UAV $m$ to UAV $m'$. The number of transmitted packets is limited by the intended allocation, link capacity, and available buffer at the receiving UAV,
\begin{align}
q_{m,m'}^{\rm trans}(t)
\leq \min\{&
g_{m,m'}(t),q_{m'}^{\rm free}(t),q_{m,m'}^{\rm ori}(t)\},
\nonumber\\
&\forall m\in{\cal M},\,m'\in\psi_m^{\rm com}(t),\,t\in{\cal T},
\label{UAVs-trans-packs}
\end{align}
where $q_{m,m'}^{\rm ori}(t)=|{\cal W}_{m,m'}^{\rm ori}(t)|$ and $q_{m,m'}^{\rm trans}(t)=|{\cal W}_{m,m'}^{\rm trans}(t)|$.\par

Since multiple upstream UAVs may transmit to the same UAV simultaneously, their total incoming traffic must also satisfy the receiver-buffer constraint. The incoming-neighbor set of UAV $m$ is defined as $\zeta^{\rm in}_m(t) = \{\, m' \in \mathcal{M} \setminus \{m\} \mid m \in \psi_{m'}^{\rm com}(t)\}$. Therefore,
\begin{equation}
\sum\nolimits_{m'\in\zeta_m^{\rm in}(t)}
q_{m',m}^{\rm trans}(t)
\leq q_m^{\rm free}(t).
\label{UAVs-recei-packs}
\end{equation}

If some packets assigned to a next-hop UAV cannot be successfully transmitted and admitted by the receiver, they are regarded as dropped. The set of packets dropped by UAV $m$ in time slot $t$ is
${\cal W}_m^{\rm loss}(t)=
\bigcup_{m'\in\psi_m^{\rm com}(t)}
\left(
{\cal W}_{m,m'}^{\rm ori}(t)
\setminus
{\cal W}_{m,m'}^{\rm trans}(t)
\right)$.\par

For a fair evaluation, we only consider traffic flows whose deadlines fall within the flight duration, since the delivery outcomes of later-deadline flows cannot be fully observed before the simulation ends. The evaluated flow set is ${\cal U}_{\rm eval}=\{u\in{\cal U}\mid t_u^{\rm end}\leq T\}$, and the corresponding packet set is ${\cal W}_{\rm eval}=\bigcup_{u\in{\cal U}_{\rm eval}}{\cal W}_u$. The set of dropped packets in ${\cal W}_{\rm eval}$ is ${\cal W}_{\rm eval}^{\rm loss}={\cal W}_{\rm eval}\cap\left(\bigcup_{t\in{\cal T}}\bigcup_{m\in{\cal M}}{\cal W}_m^{\rm loss}(t)\right)$. The overall packet loss ratio is
\begin{equation}
\varphi_{\rm loss}
=
\frac{|{\cal W}_{\rm eval}^{\rm loss}|}
{|{\cal W}_{\rm eval}|}.
\label{loss-ratio}
\end{equation}

Let $t_w^{\rm arr}$ denote the arrival time of packet $w\in{\cal W}$ at the GBS. For a packet that never reaches the GBS, we set $t_w^{\rm arr}=+\infty$. The set of packets delivered no later than their deadlines is ${\cal W}_{\rm eval}^{\rm on}=\{w\in{\cal W}_{\rm eval}\mid t_w^{\rm arr}\leq t_w^{\rm end}\}$. The on-time packet delivery ratio is then
\begin{equation}
\eta_{\rm pack}
=
\frac{|{\cal W}_{\rm eval}^{\rm on}|}
{|{\cal W}_{\rm eval}|}.
\label{pack-deli-ratio}
\end{equation}

\subsection{Problem Formulation}

This work aims to develop a traffic-adaptive per-hop multipath routing strategy that improves the on-time packet delivery ratio while reducing the packet loss ratio. Let ${\bf A}=\{{\bm a}_m(t)\mid m\in{\cal M},\,t\in{\cal T}\}$ denote the routing decisions of all UAVs over the flight duration. The optimization problem is formulated as\phantomsection
\label{P1}
\begin{align}
{\rm (P1):}\quad
&\mathop{\max}\limits_{\bf A}\quad
\eta_{\rm pack}-\varphi_{\rm loss}
\nonumber\\
&{\rm s.t.}\quad
\eqref{time-slot-constraint},\ \eqref{forward-0-1}-\eqref{UAVs-recei-packs}.
\nonumber
\end{align}Constraint \eqref{time-slot-constraint} specifies the maximum slot duration. Constraints \eqref{forward-0-1} and \eqref{forward-sum} ensure that the traffic-splitting ratios lie within $[0,1]$ and sum to one. Constraint \eqref{UAVs-trans-packs} limits the number of packets transmitted to each next-hop UAV according to the link capacity, packet allocation, and available receiver buffer. Constraint \eqref{UAVs-recei-packs} further ensures that the total packets received by each UAV do not exceed its available buffer capacity.

\section{Problem Solution}
Problem (\hyperref[P1]{P1}) involves sequential decision-making among multiple UAVs under partial observability in a dynamic network environment. To address this problem, we first formulate it as a Dec-POMDP and then develop the MAPPO-DM algorithm.

\subsection{Dec-POMDP Formulation}

We formulate problem~(\hyperref[P1]{P1}) as a Dec-POMDP, $\mathcal{D}=\left\langle
{\cal M},
{\cal S},
\{{\cal O}_m\}_{m\in{\cal M}},
\{{\cal A}_m\}_{m\in{\cal M}},
{\cal P},
{\cal R}
\right\rangle$. Here, ${\cal M}$ is the set of UAV agents, ${\cal S}$ is the global state space, and ${\cal O}_m$ and ${\cal A}_m$ are the local observation and action spaces of UAV $m$, respectively. ${\cal P}$ denotes the state-transition function, and ${\cal R}$ denotes the shared reward function.

In time slot $t$, each UAV $m$ observes ${\bm o}_m(t)\in{\cal O}_m$ and selects a traffic-splitting action ${\bm a}_m(t)\in{\cal A}_m$. After all UAVs execute their actions, the network transitions from ${\bm s}(t)\in{\cal S}$ to ${\bm s}(t+1)$ according to ${\cal P}$, and the UAVs receive a shared immediate reward determined by ${\cal R}$.\par

\emph{1) Observation Space:} At each time slot, the local observation of UAV $m$ consists of the traffic information and candidate-neighbor information, given by ${\bm o}_m(t)=\left[{\bm o}^{\rm traf}_m(t),{\bm o}^{\rm nei}_m(t)\right]$, where ${\bm o}^{\rm traf}_m(t)$ characterizes the currently selected traffic, and ${\bm o}^{\rm nei}_m(t)$ contains the features of the candidate next-hop UAVs.\par

The flow information consists of the number of selected packets and their delay urgency. Since traffic splitting is performed for the highest-priority nonempty queue, its urgency is determined by the earliest packet deadline. The minimum remaining time and the corresponding urgency are defined as $\tau_m^{\rm rem}(t)={\rm max}\{0,{\rm min}_{w\in Q_m^{\rm sel}(t)}(t_w^{\rm end}-t\Delta t)\}$ and $\iota_m^{\rm urg}(t)=1-{\rm min}\{1,\tau_m^{\rm rem}(t)/\tau^{\rm ref}\}$, respectively, where $\tau^{\rm ref}$ is the normalization reference. A larger $\iota_m^{\rm urg}(t)$ indicates greater forwarding urgency. Thus, the traffic feature vector is expressed as
\begin{equation}
{\bm o}^{\rm traf}_m(t)
=[q_m^{\rm sel}(t),
\iota_m^{\rm urg}(t)].
\label{flow-information}
\end{equation}

Due to UAV mobility, the candidate next-hop set changes dynamically over time. Let $K_m(t)=|\psi_m^{\rm com}(t)|$ denote the number of candidate next hops of UAV $m$ at time slot $t$. To maintain a fixed and explicit correspondence between neighbor features and action components, UAV $m$ sorts its candidate next hops in ascending order of their global identifiers, yielding ${\bm \sigma}_m(t)=\left[\sigma_m^1(t),\ldots,\sigma_m^{K_m(t)}(t)\right]$, where $\sigma_m^n(t), 1 \leq n \leq K_m(t)$ denotes the $n$-th candidate next hop after sorting.

For the $n$-th candidate next hop $m'=\sigma_m^n(t)$, the normalized geographical progress is defined as $p_{m,m'}^{\rm geo}(t)=({d_{m,k}(t)-d_{m',k}(t)})/{d^{\rm ref}}$, where $d^{\rm ref}$ is the normalization reference. This metric quantifies the progress made toward the GBS when packets are forwarded through UAV $m'$.\par

The neighbor features should also capture the communication quality of each candidate link. Although the receiver can measure the current co-channel interference, collecting and feeding back such information before routing would introduce additional signaling delay. To enable immediate forwarding after the initial information exchange at the beginning of each time slot, the current interference and link capacity are approximated from historical receiver-side measurements using an exponentially weighted moving average (EWMA).\par

Specifically, at the end of each time slot, each UAV records the aggregate interference measured on its active subbands. For each subband, only the latest measurement and its time-slot index are retained. Let $t_{j,b}^{\rm rec}$ denote the most recent time slot in which UAV $j$ measured the aggregate interference on subband $b\in{\cal B}$, and let $\ell_{j,b}(t_{j,b}^{\rm rec})$ denote the corresponding measurement. The EWMA interference estimate of UAV $j$ on subband $b$ at time slot $t$ is defined as $\bar{\ell}_{j,b}(t)=\beta_{\rm ew}\bar{\ell}_{j,b}(t-1)+(1-\beta_{\rm ew})\ell_{j,b}(t_{j,b}^{\rm rec})$, where $\beta_{\rm ew}\in[0,1]$ is the smoothing factor. Moreover, since dynamic topology and subband allocation may render some historical measurements outdated, a default interference level is introduced. The average of the latest interference measurements across all subbands is defined as $\ell_j^{\rm avg}(t)=\frac{1}{B}\sum_{b=1}^{B}\ell_{j,b}(t_{j,b}^{\rm rec})$, and the default interference estimate is updated as ${\ell}_j^{\rm base}(t)=\beta_{\rm ew}{\ell}_j^{\rm base}(t-1)+(1-\beta_{\rm ew})\ell_j^{\rm avg}(t)$. The freshness weight is defined as $\beta_{j,b}^{\rm dyn}(t)=\exp(-\max\{0,t-t_{j,b}^{\rm rec}-1\}/\tau_{\rm hist})$, where $\tau_{\rm hist}$ controls the decay rate of outdated interference information. The effective interference estimate is then given by $\hat{\ell}_{j,b}(t)=\beta_{j,b}^{\rm dyn}(t)\bar{\ell}_{j,b}(t)+(1-\beta_{j,b}^{\rm dyn}(t)){\ell}_j^{\rm base}(t)$. When $t-t_{j,b}^{\rm rec}=1$, $\beta_{j,b}^{\rm dyn}(t)=1$, and the latest subband-specific estimate is directly used. As the measurement becomes outdated, $\beta_{j,b}^{\rm dyn}(t)$ decreases, causing the estimate to gradually revert to the default interference level.\par

If UAV $m$ transmits packets to the candidate next-hop UAV $m'$ over subband $b\in{\cal B}$ at time slot $t$, the receiver-side interference estimate $\hat{\ell}_{m',b}(t)$ is used to compute the link rate and packet transmission capacity. The resulting estimated communication capacity is denoted by $\hat g_{m,m'}(t)$.\par

For the $n$-th candidate next hop $m'=\sigma_m^n(t)$, let $o_{m,n}^{\rm valid}(t) \in \{0,1\}$ denote its validity indicator. Specifically, $o_{m,n}^{\rm valid}(t)=1$ for $n\leq K_m(t)$ and $o_{m,n}^{\rm valid}(t)=0$ otherwise. The capacity-demand matching degree is further defined as $c_{m,m'}^{\rm est}(t)={\hat g_{m,m'}(t)}/{\max\{1,q_m^{\rm sel}(t)}\}$. A larger value indicates that the candidate link is better able to accommodate the current forwarding demand.\par

In addition to the geographical progress $p_{m,m'}^{\rm geo}(t)$, validity indicator $o_{m,n}^{\rm valid}(t)$, and capacity-demand matching degree $c_{m,m'}^{\rm est}(t)$, the candidate-neighbor features also include the available buffer space at the receiver $q_{m'}^{\rm free}(t)$ and the number of its upstream neighbors $|\zeta_{m'}^{\rm in}(t)|$. The latter indicates how many UAVs may simultaneously transmit packets to UAV $m'$ and thus captures its potential congestion risk. The feature vector of the $n$-th candidate neighbor is constructed as
\begin{align}
{\bm o}_{m,n}^{\rm nei}(t) =
[ 
& o_{m,n}^{\rm valid}(t),
p_{m,m'}^{\rm geo}(t), \nonumber \\
& c_{m,m'}^{\rm est}(t),  
q_{m'}^{\rm free}(t),
|\zeta_{m'}^{\rm in}(t)|
].
\label{neighbor-token}
\end{align}

Following the neighbor order in ${\bm \sigma}_m(t)$, the neighbor observation of UAV $m$ is represented as ${\bm o}_m^{\rm nei}(t)=[{\bm o}_{m,1}^{\rm nei}(t),\ldots,{\bm o}_{m,N}^{\rm nei}(t)]$. When $K_m(t)<N$, the remaining feature vectors from $K_m(t)+1$ to $N$ are zero-padded.\par

\emph{2) State Space:}
The global network state at time slot $t$ is represented as a graph. Let ${\cal V}={\cal M}$ denote the node set, where each node corresponds to a UAV. The forward-edge set is defined as ${\cal E}^{\rm fwd}(t)=\{(m,m')\mid m\in{\cal M},m'\in\psi_m^{\rm com}(t)\}$, where each edge follows the actual packet-forwarding direction. In addition, the reverse-edge and self-loop sets are defined as ${\cal E}^{\rm rev}(t)=\{(m',m)\mid(m,m')\in{\cal E}^{\rm fwd}(t)\}$ and ${\cal E}^{\rm self}=\{(m,m)\mid m\in{\cal M}\}$, respectively. The complete edge set is therefore ${\cal E}(t)={\cal E}^{\rm fwd}(t)\cup{\cal E}^{\rm rev}(t)\cup{\cal E}^{\rm self}$. Accordingly, the global state is represented as
\begin{equation}
{\cal G}(t)=
(
{\cal V},
{\cal E}(t),
{\bm X}(t),
{\bm E}(t)
),
\label{critic-graph-state}
\end{equation}
where ${\bm X}(t)=[{\bm x}_m(t)]_{m\in{\cal M}}$ and ${\bm E}(t)=[{\bm e}_{i,j}(t)]_{(i,j)\in{\cal E}(t)}$ denote the node- and edge-feature matrices, respectively.\par

For UAV $m$, let $\delta_m^{\rm GBS}(t)\in \{0,1\}$ indicate whether it can directly transmit packets to the GBS. Moreover, ${\bm x}_m^{\rm role}\in \{0,1\}^{4}$ denotes the one-hot role vector of UAV $m$, with its four entries corresponding to the hotspot, gateway, relay, and regular UAV roles, respectively. Its queue occupancy is defined as $q_m^{\rm occ}(t)=q_m^{\max}-q_m^{\rm free}(t)$. The node features further include the available buffer space, flow urgency, number of selected packets, numbers of candidate downstream and upstream neighbors, and traffic-generation probability. Thus, the node feature vector is given by
\begin{align}
{\bm x}_m(t)=
[
&q_m^{\rm occ}(t),
q_m^{\rm free}(t),
\iota_m^{\rm urg}(t),
q_m^{\rm sel}(t), \nonumber \\
&|\psi_m^{\rm com}(t)|, 
|\zeta_m^{\rm in}(t)|,
\delta_m^{\rm GBS}(t),
p_m^{\rm gen},
{\bm x}_m^{\rm role}
].
\label{critic-node-feature}
\end{align}Here, $p_m^{\rm gen}\in \{p_{\rm hot}^{\rm gen},p_{\rm gat}^{\rm gen},p_{\rm rel}^{\rm gen},p_{\rm reg}^{\rm gen}\}$ denotes the traffic-generation probability of UAV $m$, whose value is determined by its role.

For each directed edge, define the direction indicator $\delta_{i,j}^{\rm dir}\in \{1,-1,0\}$, where $\delta_{i,j}^{\rm dir}=1$ for $(i,j)\in{\cal E}^{\rm fwd}(t)$, $\delta_{i,j}^{\rm dir}=-1$ for $(i,j)\in{\cal E}^{\rm rev}(t)$, and $\delta_{i,j}^{\rm dir}=0$ for $i=j$. For a forward edge $(m,m')\in{\cal E}^{\rm fwd}(t)$, the edge features comprise the direction indicator, geographical progress, estimated and actual communication capacities, receiver-side available buffer space, and actual SINR,
\begin{align}
{\bm e}_{m,m'}(t)
=
[
& \delta_{m,m'}^{\rm dir},
p_{m,m'}^{\rm geo}(t),
\hat g_{m,m'}(t), \nonumber \\
& g_{m,m'}(t),
q_{m'}^{\rm free}(t),
\vartheta_{m,m'}(t)
].
\label{critic-edge-feature}
\end{align}
For the corresponding reverse edge $(m',m)$, $\delta_{m',m}^{\rm dir}$ is set to $-1$, while the remaining features are inherited from the forward edge $(m,m')$. All features of a self-loop edge $(m,m)$ are set to zero.\par

To improve numerical stability during training, all continuous features are normalized according to their physical ranges and clipped when necessary.

\emph{3) Action Space:} At time slot $t$, UAV $m$ forwards packets from its sub-queue $Q_m^{\text{sel}}(t)$ either to its neighbors or retains them locally. This decision is represented by the action vector
\begin{equation}
    {\bm a}_m(t) = [a_{m,0}(t), \, a_{m,1}(t), \ldots, a_{m,N}(t)],  \forall m,t,
    \label{action}
\end{equation}where $a_{m,0}(t)$ represents the proportion of packets retained locally, and $a_{m,n}(t)$ represents the proportion forwarded to the $n$-th neighbor in $\bm \sigma_m(t)$. These variables satisfy $0 \leq a_{m,n}(t) \leq 1$ for all $n \in \{0,1,\ldots,N\}$. The action vector must satisfy the normalization constraint
\begin{equation}
    \sum_{n=0}^{N} a_{m,n}(t) = 1, \quad \forall m,t.
    \label{action-sum}
\end{equation}

\emph{4) Reward Function:} Problem~(\hyperref[P1]{P1}) aims to improve on-time packet delivery while reducing packet loss. To provide immediate feedback for routing decisions, we design a dense reward that combines local forwarding performance with shared rewards for on-time arrivals and deadline misses.\par

The routing reward of UAV $m$ in time slot $t$ is defined as 
\begin{equation}
r_m(t)
=
\delta_m^{\rm act}(t)
\upsilon_m^{\rm urg}(t)
\sum_{n=0}^{N}a_{m,n}(t)r_{m,n}(t)
+
r^{\rm glo}(t),
\label{rout-rwd}
\end{equation}where $\delta_m^{\rm act}(t)\in\{0,1\}$ indicates whether UAV $m$ makes an active routing decision. Specifically, $\delta_m^{\rm act}(t)=1$ when UAV $m$ has packets awaiting forwarding, has at least one candidate next hop, and cannot directly transmit to the GBS. Otherwise, $\delta_m^{\rm act}(t)=0$. Moreover, $r_{m,n}(t)$ denotes the local reward associated with action component $a_{m,n}(t)$, while $r^{\rm glo}(t)$ is a shared event reward determined by network-wide on-time packet arrivals and deadline misses. The urgency-aware scaling factor is given by $\upsilon_m^{\rm urg}(t)=1+\chi_{\rm urg}\iota_m^{\rm urg}(t)$, where $\chi_{\rm urg}\geq0$ controls the influence of traffic urgency on the local routing reward.\par

For a forwarding component $a_{m,n}(t)$ with $n\neq 0$, let $m'=\sigma_m^{n}(t)$ denote the corresponding next-hop UAV. The forwarding reward consists of a geographic progress gain $r_{m,n}^{\mathrm{prog}}(t)$, a congestion-risk penalty $r_{m,n}^{\mathrm{cong}}(t)$, and a packet-loss penalty $r_{m,n}^{\mathrm{loss}}(t)$. It is defined as
\begin{equation}
r_{m,n}(t)= \upsilon^{\mathrm{prog}}r_{m,n}^{\mathrm{prog}}(t)
-\upsilon^{\mathrm{cong}}r_{m,n}^{\mathrm{cong}}(t) -\upsilon^{\mathrm{loss}}r_{m,n}^{\mathrm{loss}}(t), 
\end{equation}where $\upsilon^{\mathrm{prog}}$, $\upsilon^{\mathrm{cong}}$, and $\upsilon^{\mathrm{loss}}$ are the weights for the corresponding reward and penalty terms.\par

To evaluate the forwarding decision, we first define four related metrics. The effective forwarding capacity  is given by $c_{m,m'}(t)={\rm min}(g_{m,m'}(t), q_{m'}^{\rm free}(t))$. The capacity-satisfaction ratio is then defined as $c_{m,n}^{\mathrm{fit}}(t)={\rm min}(1,{c_{m,m'}(t)}/{{\rm max}(1,q_{m,m'}^{\mathrm{ori}}(t))})$. The actual transmission ratio and packet-loss ratio are respectively given by $y_{m,n}^{\mathrm{tx}}(t)= {\rm min}(1,{q_{m,m'}^{\mathrm{trans}}(t)}/{{\rm max}(1,q_{m,m'}^{\mathrm{ori}}(t))}),$ and $y_{m,n}^{\rm loss}(t)=1-y_{m,n}^{\rm tx}(t)$.\par

Based on these metrics, the geographic progress gain is given by $r_{m,n}^{\mathrm{prog}}(t)=c_{m,n}^{\mathrm{fit}}(t)y_{m,n}^{\mathrm{tx}}(t)p_{m,n}^{\mathrm{geo}}(t)$. It rewards forwarding decisions that make progress toward the GBS while providing sufficient capacity and successfully delivering packets to the next hop. The congestion-risk penalty is defined as $r_{m,n}^{\mathrm{cong}}(t)=1-c_{m,n}^{\mathrm{fit}}(t),$ discouraging links that cannot adequately support the intended forwarding load. Finally, the packet-loss penalty is given by $r_{m,n}^{\mathrm{loss}}(t)=y_{m,n}^{\mathrm{loss}}(t)+\upsilon_{\mathrm{loss}}^{\mathrm{quad}}\left(y_{m,n}^{\mathrm{loss}}(t)\right)^2,$ where $\upsilon_{\mathrm{loss}}^{\mathrm{quad}}$ controls the quadratic penalty and imposes a sharper penalty at high packet-loss ratios.\par

For the local holding action $a_{m,0}(t)$, the reward distinguishes between necessary and unnecessary packet holding based on the aggregate forwarding capacity of all candidate next hops. Specifically, the aggregate downstream capacity is defined as $c_m^{\rm agg}(t)=\sum_{n=1}^{K_m(t)}c_{m,\sigma_m^n(t)}(t)$. If $c_m^{\rm agg}(t)$ is sufficient to accommodate the selected traffic $q_m^{\rm sel}(t)$, holding the packets is considered unnecessary and is penalized; otherwise, no holding penalty is imposed. Accordingly,
\begin{equation}
r_{m,0}(t)=
\begin{cases}
-\upsilon_m^{\mathrm{hold}}(t),& c_m^{\rm agg}(t)\geq q_m^{\rm sel}(t),\\
0,&c_m^{\rm agg}(t)<q_m^{\rm sel}(t),
\end{cases}
\label{hold-pena}
\end{equation}
where $\upsilon_m^{\mathrm{hold}}(t)=K_0^{\mathrm{hold}}+K_1^{\mathrm{hold}}\iota_m^{\mathrm{urg}}(t)$. Here, $K_0^{\mathrm{hold}}$ is the basic holding-penalty coefficient, while $K_1^{\mathrm{hold}}$ increases the penalty for more urgent traffic.\par

To directly incorporate end-to-end packet-delivery outcomes into the reward, a network-wide event reward is further introduced. Let $W^{\rm on}(t)$ denote the number of packets that arrive at the GBS on time during time slot $t$, and let $W^{\rm mis}(t)$ denote the number of packets that miss their deadlines during the same time slot. Moreover, let $\delta^{\rm all}(t) = \sum_{m\in{\cal M}}\delta_m^{\rm act}(t)$ denote the number of valid UAVs. The global event reward assigned to UAV $m$ is given by
\begin{align}
r_m^{\rm glo}(t)
=
\delta_m^{\rm act}(t) (&{\upsilon_{\rm on}^{\rm glo} W^{\rm on}(t)} / {\max\{1,\delta^{\rm all}(t)} \}, \nonumber \\
& - {\upsilon_{\rm mis}^{\rm glo}W^{\rm mis}(t)} / {\max\{1,\delta^{\rm all}(t)} \}),
\end{align} where $\upsilon_{\rm on}^{\rm glo}$ and $\upsilon_{\rm mis}^{\rm glo}$ control the contributions of on-time arrivals and deadline misses. When $\delta^{\rm all}(t) >0$, the global event reward and penalty are equally distributed among the valid UAVs.

\subsection{The Proposed MAPPO-DM Algorithm}

Based on the Dec-POMDP formulation, we develop a MAPPO-DM algorithm. The proposed algorithm follows the centralized-training-and-decentralized-execution paradigm. 

\begin{figure}
\centering
\includegraphics[width=0.75\linewidth]{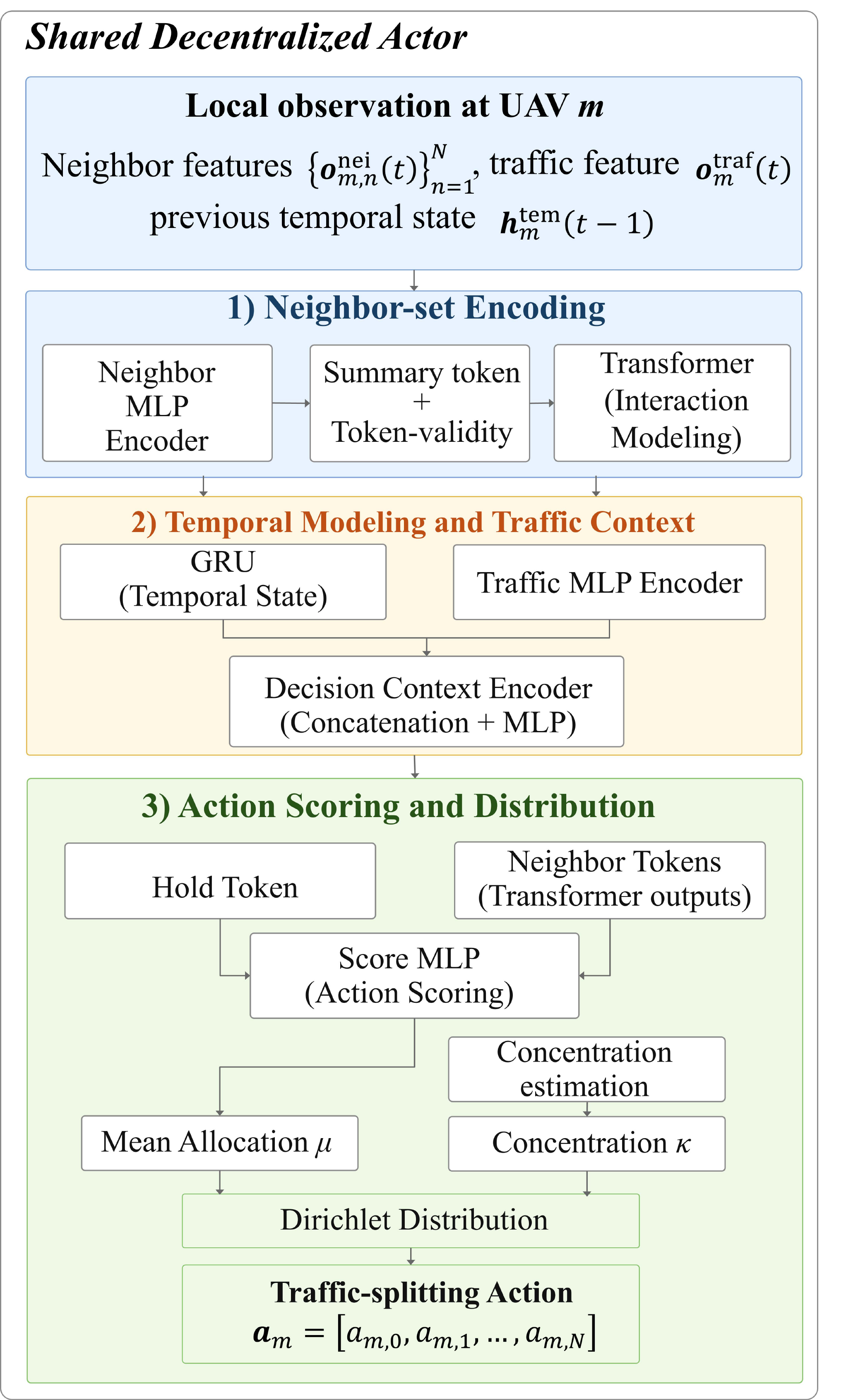}
\caption{Architecture of the Actor network.}
\label{actor}
\end{figure}

\emph{1) Decentralized Actor Design:}
The decentralized policy of UAV $m$ is denoted by
$\pi_{\theta_{\rm A}}$, where $\theta_{\rm A}$ represents the Actor parameters shared among all UAV agents. The Actor combines neighbor-set encoding, recurrent temporal modeling, traffic-conditioned action scoring, and Dirichlet concentration estimation. \par

For each candidate-neighbor position $n\in\{1,\ldots,N\}$, the corresponding feature vector ${\bm o}^{\rm nei}_{m,n}(t)$ is encoded into a $d_{\rm enc}^{\rm nei}$-dimensional token,
\begin{equation}
{\bm h}^{\rm nei}_{m,n}(t)
=
f_{\rm nei}
\left(
{\bm o}^{\rm nei}_{m,n}(t)
\right),
\label{actor-neighbor-embedding}
\end{equation}where $f_{\rm nei}(\cdot)$ is a multilayer perceptron (MLP).\par

To model the interactions among candidate next hops, a learnable summary token ${\bm h}_{\rm sum}\in\mathbb R^{d_{\rm enc}^{\rm nei}}$ is prepended to the neighbor tokens,
\begin{equation}
{\bm H}_m(t)
=
[
{\bm h}_{\rm sum},
{\bm h}^{\rm nei}_{m,1}(t),
\ldots,
{\bm h}^{\rm nei}_{m,N}(t)
].
\label{actor-transformer-input}
\end{equation}The corresponding token-validity vector is defined as ${\bm v}_m(t)=[1,o_{m,1}^{\rm valid}(t),\ldots,o_{m,N}^{\rm valid}(t)]$, where the first entry indicates that the summary token is always valid. The tokens are then processed by a Transformer encoder ${\rm Tr}(\cdot)$~\cite{transformer-Vaswani},
\begin{equation}
[
\widetilde{\bm h}^{\rm sum}_m(t),
\widetilde{\bm h}^{\rm nei}_{m,1}(t),
\ldots,
\widetilde{\bm h}^{\rm nei}_{m,N}(t)
]
=
{\rm Tr}
(
{\bm H}_m(t),
{\bm v}_m(t)
),
\label{actor-set-transformer}
\end{equation}where positions with $o_{m,n}^{\rm valid}(t)=0$ are masked during self-attention. Since the identifier-based neighbor order is used only to align neighbor features with the corresponding action components, no positional encoding is introduced.\par

The Transformer captures the interactions among candidate neighbors within each time slot but does not retain information across time slots. To capture the temporal evolution of the local network state, the contextualized neighbor-set summary is used to update a recurrent hidden state,
\begin{equation}
{\bm h}^{\rm tem}_m(t)
=
{\rm GRU}
\left(
f_{\rm proj}
\left(
\widetilde{\bm h}^{\rm sum}_m(t)
\right),
{\bm h}^{\rm tem}_m(t-1)
\right),
\label{actor-gru}
\end{equation}where ${\rm GRU}(\cdot)$ denotes a GRU that updates the current temporal hidden state based on the projected neighbor-set summary and the hidden state from the previous time slot. The MLP $f_{\rm proj}(\cdot)$ maps $\widetilde{\bm h}^{\rm sum}_m(t)$ to the input space of the GRU.\par

Meanwhile, the traffic features are encoded as
\begin{equation}
{\bm h}^{\rm traf}_m(t)
=
f_{\rm traf}
\left(
{\bm o}^{\rm traf}_m(t)
\right),
\label{actor-flow-embedding}
\end{equation}
where $f_{\rm traf}(\cdot)$ is an MLP-based encoder that maps the traffic features into a latent representation. The temporal hidden state and traffic embedding are then concatenated to form the decision context,
\begin{equation}
{\bm h}_m^{\rm ctx}(t)
=
{\rm concat}
\left(
{\bm h}^{\rm tem}_m(t),
{\bm h}^{\rm traf}_m(t)
\right).
\end{equation}
The resulting representation is further transformed into an action-scoring context,
\begin{equation}
{\bm h}^{\rm acs}_m(t)
=
f_{\rm acs}
\left(
{\bm h}_m^{\rm ctx}(t)
\right),
\label{actor-decision-context}
\end{equation}where $f_{\rm acs}(\cdot)$ is an MLP-based context encoder that extracts the latent features used for subsequent action scoring.\par

To explicitly represent the local holding action, a learnable holding token ${\bm h}^{\rm hold}\in\mathbb R^{d_{\rm enc}^{\rm nei}}$ is introduced. The token associated with action component $n\in\{0,\ldots,N\}$ is defined as
\begin{equation}
{\bm a}^{\rm tok}_{m,n}(t)
=
\begin{cases}
{\bm h}^{\rm hold}, & n=0,\\
\widetilde{\bm h}^{\rm nei}_{m,n}(t), & 1\leq n\leq N.
\end{cases}
\label{actor-action-token}
\end{equation}
Each action token is then concatenated with the action-scoring context and mapped to a scalar score,
\begin{equation}
\lambda_{m,n}(t)
=
f_{\rm score}
\left(
{\rm concat}
\left(
{\bm a}^{\rm tok}_{m,n}(t),
{\bm h}^{\rm acs}_m(t)
\right)
\right),
\label{actor-action-score}
\end{equation}
where $f_{\rm score}(\cdot)$ is an MLP-based scoring network. It evaluates each candidate action conditioned on the current decision context and outputs the corresponding unnormalized score.\par

Let $o_{m,0}^{\rm valid}(t)=1$, since the holding action is always available. A masked softmax produces the mean traffic-splitting preference,
\begin{equation}
\mu_{m,n}(t)
=
\frac{
o_{m,n}^{\rm valid}(t)\exp(\lambda_{m,n}(t))
}{
\sum_{j=0}^{N}
o_{m,j}^{\rm valid}(t)\exp(\lambda_{m,j}(t))
},
\label{actor-mean-action}
\end{equation}Thus, invalid neighbor positions receive zero probability, while the valid components satisfy $\sum_{n=0}^{N}\mu_{m,n}(t)=1$.

Since the routing action is a continuous traffic-splitting vector defined on the probability simplex, the Actor parameterizes its stochastic policy using a Dirichlet distribution~\cite{dirichlet-distri}. Let ${\bm \mu}_m(t)=[\mu_{m,0}(t),\ldots,\mu_{m,N}(t)]$ denote the mean traffic-splitting vector. To separately control the stochasticity of the policy, the decision context ${\bm h}_m(t)$ is further used to estimate the total concentration parameter,
\begin{equation}
\kappa_m(t)
=
\min
\left\{
\kappa_{\max},
\kappa_{\min}
+
\operatorname{softplus}
\left(
f_{\kappa}
\left(
{\bm h}_m^{\rm ctx}(t)
\right)
\right)
\right\},
\label{actor-kappa}
\end{equation}where $f_{\kappa}(\cdot)$ is an MLP-based concentration estimator with a scalar output. The lower and upper bounds $\kappa_{\min}$ and $\kappa_{\max}$ constrain the total concentration to a numerically stable range.\par

The Dirichlet concentration parameter of action component $n$ is
then constructed as
\begin{equation}
\alpha_{m,n}(t)
=
\kappa_m(t)\mu_{m,n}(t).
\label{actor-dirichlet-alpha}
\end{equation}
Thus, ${\bm \mu}_m(t)$ determines the mean traffic allocation,
whereas $\kappa_m(t)$ controls how strongly the sampled action is
concentrated around this mean. Specifically, A larger $\kappa_m(t)$ produces more stable traffic splitting,
whereas a smaller value encourages stronger stochastic exploration.\par

Let $\mathcal{I}_m(t)=\{0\}{\cup}\{n\in\{1,...,N\}| o_{m,n}^{\rm valid}(t)=1\}$ denote the set of valid action indices for UAV $m$. The corresponding concentration and action subvectors are defined as $\bm{\alpha}_m^{\rm val}(t)=[\alpha_{m,n}(t)]_{n\in\mathcal{I}_m(t)}$ and $\bm{a}_m^{\rm val}(t)=[a_{m,n}(t)]_{n\in\mathcal{I}_m(t)}$, respectively. The valid action components are then sampled on the corresponding probability simplex according to
\begin{equation}
\bm{a}_m^{\rm val}(t)
\sim
\operatorname{Dir}
\left(
\bm{\alpha}_m^{\rm val}(t)
\right),
\label{actor-dirichlet-sampling}
\end{equation}while the components associated with invalid padded positions are set to zero.\par

During training, the sampled action is used for stochastic exploration and PPO policy evaluation, whereas the mean traffic-splitting vector ${\bm \mu}_m(t)$ is used for deterministic execution.\par

\begin{figure}
\centering
\includegraphics[width=0.75\linewidth]{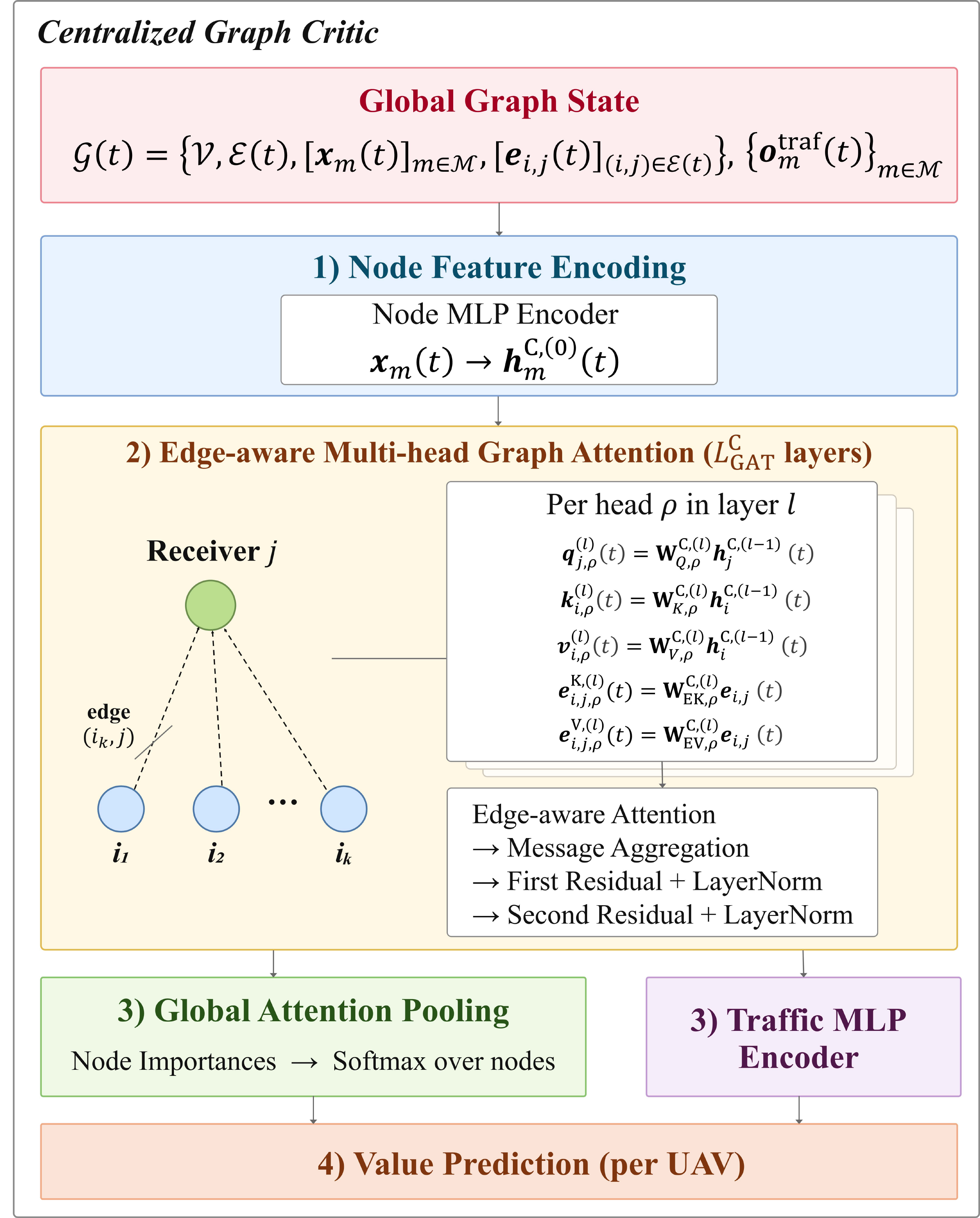}
\caption{Architecture of the Critic network.}
\label{critic}
\end{figure}

\emph{2) Centralized Critic Design:} The centralized Critic is denoted by $V_{\theta_{\rm C}}$, where $\theta_{\rm C}$ represents its trainable parameters. During training, the Critic takes the global graph state ${\mathcal G}(t)$ and the traffic features $\{{\bm o}^{\rm traf}_m(t)\}_{m \in \cal M}$ as inputs to estimate an individual state value for each UAV. It consists of five main components, namely node feature encoding, edge-aware multi-head graph attention, global attention pooling, traffic feature encoding, and a value prediction head.\par

First, the node feature $ {\bm x}_m(t)$ of UAV $m$ is mapped into a $d_{\mathrm{hid}}^{\mathrm C}$-dimensional hidden space to obtain its initial node representation, 
\begin{equation}
\bm{h}_m^{\mathrm C,(0)}(t)
=
f_{\mathrm{node}}^{\mathrm C}
\left(
\bm{x}_m(t)
\right),
\end{equation}where $f_{\mathrm{node}}^{\mathrm C}(\cdot)$ is an MLP-based node encoder. The resulting $\bm{h}_m^{\mathrm C,(0)}(t)$ serves as the input to the first graph attention layer.\par

Following the node encoding, the Critic employs an $L_{\mathrm{GAT}}^{\mathrm C}$-layer edge-aware multi-head graph attention network. Each layer contains $n_{\mathrm{head}}^{\mathrm C}$ attention heads, each with dimension $d_{\mathrm{head}}^{\mathrm C}=d_{\mathrm{hid}}^{\mathrm C}/n_{\mathrm{head}}^{\mathrm C}$.\par

Specifically, for a directed edge $(i,j)\in\mathcal{E}(t)$, nodes $i$ and $j$ are the message sender and receiver, respectively. In the $\rho$-th attention head of the $l$-th layer, the receiver representation is projected into a Query, while the sender representation is projected into a Key and a Value. They are given by
\begin{align}
\bm{q}_{j,\rho}^{(l)}(t)
&=
\mathbf{W}_{{\rm Q},\rho}^{\mathrm C,(l)}
\bm{h}_j^{\mathrm C,(l-1)}(t),\\
\bm{k}_{i,\rho}^{(l)}(t)
&=
\mathbf{W}_{{\rm K},\rho}^{\mathrm C,(l)}
\bm{h}_i^{\mathrm C,(l-1)}(t),\\
\bm{v}_{i,\rho}^{(l)}(t)
&=
\mathbf{W}_{{\rm V},\rho}^{\mathrm C,(l)}
\bm{h}_i^{\mathrm C,(l-1)}(t).
\end{align}Here, $\mathbf{W}_{{\rm Q},\rho}^{\mathrm C,(l)}$, $\mathbf{W}_{{\rm K},\rho}^{\mathrm C,(l)}$, and $\mathbf{W}_{{\rm V},\rho}^{\mathrm C,(l)}$ are the corresponding trainable projection matrices. \par

Meanwhile, the edge feature ${\bm e}_{i,j}(t)$ is projected into an edge-key and an edge-value as
\begin{align}
\bm{e}_{i,j,\rho}^{{\rm K},(l)}(t)
&=
\mathbf{W}_{{\rm EK},\rho}^{\mathrm C,(l)}
\bm{e}_{i,j}(t),\\
\bm{e}_{i,j,\rho}^{{\rm V},(l)}(t)
&=
\mathbf{W}_{{\rm EV},\rho}^{\mathrm C,(l)}
\bm{e}_{i,j}(t),
\end{align}where $\mathbf{W}_{{\rm EK},\rho}^{\mathrm C,(l)}$ and $\mathbf{W}_{{\rm EV},\rho}^{\mathrm C,(l)}$ are trainable edge-feature projection matrices.\par

Next, define the incoming-neighbor set of node $j$ as $\mathcal{N}_{j}^{\mathrm{in}}(t)=\{i\mid(i,j)\in\mathcal{E}(t)\}$. For each $i\in\mathcal{N}_{j}^{\mathrm{in}}(t)$, the Key of sending node $i$ is combined with the corresponding edge-key and matched with the Query of receiving node $j$. The resulting attention score is
\begin{equation}
{\hat s}_{i,j,\rho}^{(l)}(t)
=
\frac{
(\bm{q}_{j,\rho}^{(l)}(t))^{\mathsf T}
(
\bm{k}_{i,\rho}^{(l)}(t)
+
\bm{e}_{i,j,\rho}^{{\rm K},(l)}(t)
)
}{
\sqrt{d_{\mathrm{head}}^{\mathrm C}}
}.
\end{equation}The attention scores are normalized over all incoming neighbors of node $j$ using the Softmax function~\cite{softmax-Bridle},
\begin{equation}
{s}_{i,j,\rho}^{(l)}(t)
=
{\rm Softmax}_{i\in\mathcal{N}_{j}^{\rm in}(t)}
\left(
{\hat s}_{i,j,\rho}^{(l)}(t)
\right).
\end{equation}Thus, the attention weights capture both node-state compatibility and link attributes.\par

The message propagated along edge $(i,j)$ is constructed as $\bm{m}_{i,j,\rho}^{(l)}(t)=\bm{v}_{i,\rho}^{(l)}(t)+\bm{e}_{i,j,\rho}^{{\rm V},(l)}(t)$. The $\rho$-th attention head then aggregates the incoming messages at node $j$ using their attention weights,
\begin{equation}
\bm{z}_{j,\rho}^{(l)}(t)
=
{\textstyle\sum}_{i\in\mathcal{N}_{j}^{\mathrm{in}}(t)}
{s}_{i,j,\rho}^{(l)}(t)
\bm{m}_{i,j,\rho}^{(l)}(t).
\end{equation}

The outputs of all attention heads are then concatenated as $\bm{z}_{j}^{(l)}(t)=\operatorname{concat}_{\rho=1}^{n_{\mathrm{head}}^{\mathrm C}}(\bm{z}_{j,\rho}^{(l)}(t))$, where $\operatorname{concat}(\cdot)$ denotes concatenation. The resulting multi-head representation is projected and combined with the previous-layer node representation through a residual connection and LayerNorm,
\begin{equation}
\widetilde{\bm{h}}_{j}^{\mathrm C,(l)}(t)
=
\mathrm{LN}_{1}^{\mathrm C,(l)}
(
\bm{h}_{j}^{\mathrm C,(l-1)}(t)
+
f_{\mathrm{proj}}^{\mathrm C,(l)}
(
\bm{z}_{j}^{(l)}(t)
)
),
\end{equation}where ${\rm LN}_{1}^{\mathrm C,(l)}(\cdot)$ denotes the first LayerNorm operation in the $l$-th layer, and $f_{\mathrm{proj}}^{\mathrm C,(l)}(\cdot)$ denotes the corresponding multi-head output projection.\par

Next, $\widetilde{\bm{h}}_{j}^{\mathrm C,(l)}(t)$ is processed by a feed-forward network. Its output is combined with $\widetilde{\bm{h}}_{j}^{\mathrm C,(l)}(t)$ through a second residual connection and normalization, yielding
\begin{equation}
\bm{h}_{j}^{\mathrm C,(l)}(t)
=
\mathrm{LN}_{2}^{\mathrm C,(l)}
(
\widetilde{\bm{h}}_{j}^{\mathrm C,(l)}(t)
+
f_{\mathrm{FFN}}^{\mathrm C,(l)}
(
\widetilde{\bm{h}}_{j}^{\mathrm C,(l)}(t)
)
),
\end{equation}where ${\rm LN}_{2}^{\mathrm C,(l)}(\cdot)$ denotes the second LayerNorm operation in the $l$-th layer, and $f_{\mathrm{FFN}}^{\mathrm C,(l)}(\cdot)$ denotes the corresponding feed-forward network. The resulting $\bm{h}_{j}^{\mathrm C,(l)}(t)$ is used as the input to the next graph attention layer.\par

After $L_{\mathrm{GAT}}^{\mathrm C}$ graph attention layers, each node representation incorporates multi-hop node and edge information. The Critic then applies attention pooling to obtain a global graph representation. \par

Specifically, a linear mapping $f_{\mathrm{gate}}^{\mathrm C}(\cdot)$ assigns an importance score to each UAV, and the corresponding attention weights and global representation are given by
\begin{align}
{s}_m^{\rm node}(t)
&=
{\rm Softmax}_{m\in\mathcal M}
(
f_{\mathrm{gate}}^{\mathrm C}
(
\bm{h}_m^{\mathrm C,(L_{\mathrm{GAT}}^{\mathrm C})}(t)
)
),\\
\bm{h}^{\mathrm C,\mathrm{glo}}(t)
&=
{\textstyle\sum}_{m\in\mathcal M}
{s}_m^{\rm node}(t)
\bm{h}_m^{\mathrm C,(L_{\mathrm{GAT}}^{\mathrm C})}(t).
\end{align}

In parallel, the current traffic features are encoded as $\bm{h}_m^{\mathrm C,\mathrm{traf}}(t)=f_{\mathrm{traf}}^{\mathrm C}({\bm o}_m^{\mathrm{traf}}(t))$, where $f_{\mathrm{traf}}^{\mathrm C}(\cdot)$ is a MLP-based traffic encoder.

Finally, the node, global graph, and traffic representations are concatenated and passed through an MLP-based value prediction head $f_{\mathrm{val}}^{\mathrm C}(\cdot)$ to obtain the state-value estimate for UAV $m$,
\begin{align}
V_m(t)
=
f_{\mathrm{val}}^{\mathrm C}
(
\operatorname{concat}
(&
\bm{h}_m^{\mathrm C,(L_{\mathrm{GAT}}^{\mathrm C})}(t),\nonumber \\
&\bm{h}^{\mathrm C,\mathrm{glo}}(t),
\bm{h}_m^{\mathrm C,\mathrm{traf}}(t)
)
),
\end{align}

\emph{3) Experience Collection and Policy Optimization:} At time slot $t$, UAV $m$ obtains its local observation ${\bm o}_m(t)$, from which the Actor constructs the Dirichlet concentration parameter vector ${\bm \alpha}_m(t)$. A stochastic routing action is then sampled as ${\bm a}_m(t)\sim{\rm Dir}({\bm \alpha}_m(t))$, and its log-probability is recorded as ${\rm LogP}_m^{\rm old}(t)=\log p({\bm a}_m(t)|{\bm \alpha}_m(t))$. After all UAVs select their actions, the environment executes the joint action and returns the immediate reward $r_m(t)$ for each UAV. At the end of the time slot, the corresponding experience is stored as ${\rm Mem}(t)=({\cal G}(t),\{({\bm o}_m(t),{\bm a}_m(t),{\rm LogP}_m^{\rm old}(t),\delta_m^{\rm act}(t),r_m(t))| m\in{\cal M}\})$. After completing the episode, the collected trajectory is denoted by ${\cal B}=\{{\rm Mem}(t)\mid t\in{\cal T}\}$.\par

For Actor updating, the stored observation sequence is replayed in temporal order to reconstruct the recurrent hidden states. The current Actor then recomputes the Dirichlet concentration parameters at each time slot and evaluates the stored actions under the current policy. For UAV $m$, the resulting action log-probability is denoted by ${\rm LogP}_m^{\rm new}(t)=\log p({\bm a}_m(t)\mid{\bm \alpha}_m(t))$, and the entropy of the corresponding Dirichlet distribution is denoted by $H({\rm Dir}({\bm \alpha}_m(t)))$.\par

The advantage is estimated using Generalized Advantage Estimation (GAE)~\cite{GAE}. Before the PPO updates, the Critic computes the reference state values $V_m^{\rm old}(t)$, which remain fixed during the subsequent updates. The temporal-difference error is ${\rm TDE}_m(t)=r_m(t)+\gamma V_m^{\rm old}(t+1)-V_m^{\rm old}(t)$, where $0\leq\gamma\leq1$ is the discount factor. The advantage is recursively computed as $A_m(t)={\rm TDE}_m(t)+\gamma\lambda A_m(t+1)$, where $0\leq\lambda\leq1$ controls the bias-variance tradeoff in GAE.\par

The probability ratio between the current policy and the behavior policy is computed as
\begin{equation}
\varpi_m(t)
=
\exp\left(
{\rm LogP}_m^{\rm new}(t)
-
{\rm LogP}_m^{\rm old}(t)
\right).
\label{new-old-policy-ratio}
\end{equation}
The PPO clipped surrogate objective is then defined as
\begin{align}
J_m^{\rm clip}(t)
=
\min\Big\{
&\varpi_m(t)A_m(t),\nonumber\\
&
{\rm clip}
\left(
\varpi_m(t),
1-\varepsilon_1,
1+\varepsilon_1
\right)
A_m(t)
\Big\},
\label{actor-loss-clip}
\end{align}
where $\varepsilon_1$ is the policy clipping threshold. Let $\mathcal{I}^{\rm act}=\{(m,t)\mid\delta_m^{\rm act}(t)=1\}$ denote the set of valid routing-decision samples. With entropy regularization, the Actor loss is defined as
\begin{equation}
L^{\rm actor}
=
-\frac{1}{|\mathcal{I}^{\rm act}|}
\sum_{(m,t)\in\mathcal{I}^{\rm act}}
[
J_m^{\rm clip}(t)
+
\wp H({\rm Dir}({\bm \alpha}_m(t)))
],
\label{actor-loss}
\end{equation}
where $\wp$ controls the strength of entropy regularization.\par

For Critic updating, a value clipping strategy is adopted to stabilize training. The clipped value estimate is defined as
\begin{equation}
V_m^{\rm clip}(t)
=
V_m^{\rm old}(t)
+
{\rm clip}
\left(
V_m(t)-V_m^{\rm old}(t),
-\varepsilon_2,
\varepsilon_2
\right),
\label{clip-value}
\end{equation}where $V_m(t)$ is the current Critic estimate and $\varepsilon_2$ is the value clipping threshold. The Critic is trained using the one-step temporal-difference target
\begin{equation}
\hat V_m(t)
=
r_m(t)+\gamma V_m^{\rm old}(t+1).
\label{value-target}
\end{equation}
The Critic loss is then defined as
\begin{align}
L^{\rm critic}
=
\frac{1}{|\mathcal{I}^{\rm act}|}
\sum_{(m,t)\in\mathcal{I}^{\rm act}}
\max\{
&(V_m(t)-\hat V_m(t))^2,
\nonumber\\[-2mm]
&(V_m^{\rm clip}(t)-\hat V_m(t))^2
\}.
\label{critic-loss}
\end{align}

Let $N_{\rm epi}$ denote the total number of training episodes and $N_{\rm upd}$ denote the number of PPO updates performed on each collected trajectory. In each episode, MAPPO-DM first collects a complete trajectory over $T_{\rm max}$ time slots and then repeatedly updates the Actor and Critic using the collected experience. The overall training procedure is summarized in Algorithm~\ref{alg1}. The source code for both training and testing is publicly available online.\footnote{\label{foot:code}Source code: \href{https://github.com/zhenyuzhao-research/traffic-adaptive-per-hop-multipath-routing}{GitHub repository}.}\par

\begin{algorithm}[t]
\caption{Training procedure of the MAPPO-DM algorithm}
\label{alg1}

Initialize the Actor parameters $\theta_{\rm A}$ and Critic parameters $\theta_{\rm C}$.\;
Initialize the experience buffer $\mathcal{B}$.\;

\ForE{$i=1,\ldots,N_{\rm epi}$}{
Clear the experience buffer $\mathcal{B}$.\;

\ForE{$t=1,\ldots,T_{\rm max}$}{
Obtain the local observations $\{{\bm o}_m(t)\mid m\in\mathcal{M}\}$ and global graph state ${\cal G}(t)$.\;

\ForE{$m=1,\ldots,M$}{
Construct the Dirichlet concentration parameters ${\bm \alpha}_m(t)$ using the Actor.\;
Sample ${\bm a}_m(t)\sim{\rm Dir}({\bm \alpha}_m(t))$.\;
Record ${\rm LogP}_m^{\rm old}(t)=\log p({\bm a}_m(t)\mid{\bm \alpha}_m(t))$.\;
}

Execute the routing actions $\{{\bm a}_m(t)\mid m\in\mathcal{M}\}$ and obtain the rewards $\{r_m(t)\mid m\in\mathcal{M}\}$.\;
Store ${\rm Mem}(t)$ in $\mathcal{B}$.\;
}

Compute the reference state values $V_m^{\rm old}(t)$ and advantages $A_m(t)$ from the collected trajectory.\;

\ForE{$n=1,\ldots,N_{\rm upd}$}{
Replay the observation sequence and compute the action log-probabilities and policy entropies.\;
Update $\theta_{\rm A}$ using $L^{\rm actor}$ in Eq.~\eqref{actor-loss}.\;
Update $\theta_{\rm C}$ using $L^{\rm critic}$ in Eq.~\eqref{critic-loss}.\;
}
}
\end{algorithm}

\subsection{Computational Complexity and Practical Feasibility}

Since online deployment only requires each UAV to execute the shared Actor locally, we focus on the forward-pass complexity of the Actor, which mainly consists of MLPs, a Transformer encoder, and a GRU. \par

For an MLP $f(\cdot)$ with layer dimensions $\{d_0^f,...,d_{L_f}^f\}$, its dominant computational cost is $C_f=\sum_{l=1}^{L_f}d_{l-1}^fd_l^f$~\cite{MLP-complex}. The Transformer processes $N+1$ tokens of dimension $d_{\rm enc}^{\rm nei}$. For $L_{\rm Tr}$ Transformer layers with $n_{\rm head}^{\rm Tr}$ attention heads and feed-forward dimension $d_{\rm ff}^{\rm Tr}$, where each head has dimension $d_{\rm head}^{\rm Tr}=d_{\rm enc}^{\rm nei}/n_{\rm head}^{\rm Tr}$, its dominant cost is $C_{\rm Tr}=L_{\rm Tr}[(N+1)(d_{\rm enc}^{\rm nei})^2+(N+1)^2d_{\rm enc}^{\rm nei}+(N+1)d_{\rm enc}^{\rm nei}d_{\rm ff}^{\rm Tr}]$~\cite{transformer-Vaswani}. The projection MLP $f_{\rm proj}(\cdot)$ outputs a $d_{\rm in}^{\rm GRU}$-dimensional vector, while the GRU hidden state has dimension $d_{\rm tem}$. Hence, the GRU cost per time slot is $C_{\rm GRU}=d_{\rm in}^{\rm GRU}d_{\rm tem}+d_{\rm tem}^2$~\cite{GRU-complex}.\par

Following the Actor architecture, $f_{\rm nei}(\cdot)$ is evaluated for $N$ candidate neighbors, while the Transformer, $f_{\rm proj}(\cdot)$, GRU, $f_{\rm traf}(\cdot)$, and $f_{\rm ctx}(\cdot)$ are each evaluated once. The scoring network $f_{\rm score}(\cdot)$ is evaluated for all $N+1$ action components, followed by an $\mathcal{O}(N)$ masked Softmax operation. Therefore, the online complexity for one UAV per time slot is $C_{\rm A}^{\rm online}=\mathcal{O}(NC_{f_{\rm nei}}+C_{\rm Tr}+C_{f_{\rm proj}}+C_{\rm GRU}+C_{f_{\rm traf}}+C_{f_{\rm ctx}}+(N+1)C_{f_{\rm score}}+N)$. \par

During deployment, each UAV only exchanges lightweight neighbor information, such as node identifiers, positions, queue states, interference estimates, and timestamps. Thus, the communication overhead is limited. Moreover, the time-slotted model does not require strict global synchronization, since neighbor information can be exchanged asynchronously and refreshed on demand.\par

\begin{table}[!t]
\renewcommand{\arraystretch}{1.4}
\caption{Key Environment Parameters}
\label{table2}
\centering
\begin{tabular}{|m{5.4cm}|>{\centering\arraybackslash}m{2.5cm}|}
\hline
\textbf{Parameter} & \textbf{Value} \\ \hline
Number of UAV nodes & $M=35$ \\ \hline
Duration of each time slot & $\Delta t=0.5\ {\rm s}$ \\ \hline
Episode duration & $T=60\ {\rm s}$ \\ \hline
Maximum number of candidate neighbors & $N=6$ \\ \hline
Reference channel gain & $\beta_0=-50\ {\rm dB}$ \\ \hline
Path-loss exponent & $h=2$ \\ \hline
Maximum UAV transmit power & $P_m^{\max}=30\ {\rm dBm}$ \\ \hline
Number of orthogonal sub-channels & $B=64$ \\ \hline
Bandwidth of each sub-channel & $b=5\ {\rm MHz}$ \\ \hline
Noise power spectral density & $N_0\approx-170\ {\rm dBm/Hz}$ \\ \hline
Carrier frequency & $f_c=6.2\ {\rm GHz}$ \\ \hline
\end{tabular}
\end{table}

\section{Numerical Results}

This section evaluates the proposed MAPPO-DM algorithm through simulations under different traffic loads, UAV network scales, and candidate-neighbor limits. The evaluation mainly considers the on-time packet delivery ratio and packet loss ratio to assess the reliability and robustness of different routing schemes.\par

\textit{1) System settings:}
The simulations are conducted in a three-dimensional UAV network covering a horizontal area of $1200\ {\rm m}\times1200\ {\rm m}$. In the implementation coordinate system, UAVs move within $x,y\in[-600,600]\ {\rm m}$ and $z\in[0,100]\ {\rm m}$, while the GBS is fixed at $[0,0,-100]\ {\rm m}$. The three-dimensional Gauss--Markov mobility model~\cite{UAV-mobile} is adopted, with the UAV velocity bounded between $[15,15,5]\ {\rm m/s}$ and $[50,50,20]\ {\rm m/s}$ along the three axes.\par

To emulate heterogeneous traffic demand, UAVs are assigned different traffic-generation roles. The per-slot task-generation probabilities of hotspot, regular, relay, and gateway UAVs are set to $0.085$, $0.015$, $0.0065$, and $0.0045$, respectively. Each generated task contains $1.0$--$2.0\ {\rm MB}$ of data, with a packet payload of $1500\ {\rm Bytes}$. The corresponding routing deadline varies from $8$ to $14\ {\rm s}$ according to the task size. The communication queue capacity is set to $3000$ packets for hotspot, regular, and relay UAVs, and $2000$ packets for gateway UAVs. Other key environment parameters are summarized in Table~\ref{table2}.\par

\begin{table}[!t]
\renewcommand{\arraystretch}{1.4}
\caption{Key MAPPO-DM Algorithm Parameters}
\label{table3}
\centering
\begin{tabular}{|m{5.4cm}|>{\centering\arraybackslash}m{2.5cm}|}
\hline
\textbf{Parameter} & \textbf{Value} \\ \hline
Dirichlet concentration bounds &
\makecell{$\kappa_{\min}=2,$ \\ $\kappa_{\max}=80$} \\ \hline
Discount factor and GAE coefficient &
\makecell{$\gamma=0.95,$ \\ $\lambda=0.95$} \\ \hline
PPO clipping thresholds &
\makecell{$\varepsilon_1=0.20,$ \\ $\varepsilon_2=0.20$} \\ \hline
Entropy regularization coefficient & $\wp=0.003$ \\ \hline
PPO update rounds & $N_{\rm upd}=2$ \\ \hline
Total training episodes & $N_{\rm epi}=1000$ \\ \hline
\end{tabular}
\end{table}

The MAPPO-DM algorithm is implemented using PyTorch. Both the Actor and Critic are optimized using AdamW, with an initial learning rate of $1\times10^{-5}$ and a weight decay coefficient of $1\times10^{-3}$. Cosine learning-rate annealing is adopted with a minimum learning rate of $1\times10^{-6}$. The main network and PPO hyperparameters are summarized in Table~\ref{table3}.\par

The remaining environment and algorithm settings, together with implementation details of the baseline and ablation methods, are available in the released source code~\textsuperscript{\ref{foot:code}}.\par

\textit{2) Benchmark Algorithms:} Four representative routing schemes are adopted as baseline algorithms:

\begin{itemize}
\item \textbf{Priority-Aware Greedy Single-Path:} This baseline forwards all selected traffic to one downstream neighbor. For urgent traffic ($u_{\rm prio}(t)=1$), it selects the neighbor closest to the GBS; otherwise, it selects the neighbor with the largest $\min(q_{m'}^{\rm free}(t),\widehat g_{m,m'}(t))$.

\item \textbf{Equal-Split Per-Hop Multipath:} This baseline equally splits traffic among all downstream neighbors, without considering link capacity or queue states.

\item \textbf{Capacity-Aware Per-Hop Multipath:} This baseline splits traffic according to the available capacity of candidate neighbors. For neighbor $m'$, define ${\hat c}_{m,m'}(t)=\min\left(1,\frac{\min(q_{m'}^{\rm free}(t),\widehat g_{m,m'}(t))}{q_m^{\rm sel}(t)}\right)$ and ${\tilde c}_m(t)=\sum_{m'}{\hat c}_{m,m'}(t)$. The total forwarding ratio is ${\hat c}^{\rm tot}_m(t)=\min(1,{\hat c}^{\rm tot}_m(t))$. Accordingly, $a_{m,0}(t)=1-{\hat c}^{\rm tot}_m(t)$ and $a_{m,m'}(t)={\hat c}^{\rm tot}_m(t){\hat c}_{m,m'}(t)/{\tilde c}_m(t)$ for ${\tilde c}_m(t)>0$.

\item \textbf{I-AOMDV-Guided Per-Hop Multipath:} This baseline is adapted from I-AOMDV~\cite{I-AOMDV}. Each UAV updates its best $K=3$ QoS-aware paths every $10$ time slots based on hop count, available bandwidth, and path stability. The path scores are mapped to next-hop weights and normalized to obtain the per-hop traffic-splitting ratios.

\end{itemize}

To reduce the effect of randomness, all algorithm performance comparisons reported below are averaged over $50$ independent runs for each environment setting.

\subsection{Performance Evaluation}

To evaluate the contribution of the key components in MAPPO-DM, we consider three ablation variants. The {w/o Graph Critic} variant replaces the centralized graph-based Critic with decentralized MLP-based Critics. The {w/o Transformer} variant replaces the Transformer-based neighbor interaction module with an MLP-based representation. The {w/o $\mu$-$\kappa$ Dirichlet} variant removes adaptive concentration estimation and fixes $\kappa$ at 30.\par

\begin{figure}
    \centering
    \includegraphics[width=0.75\linewidth]{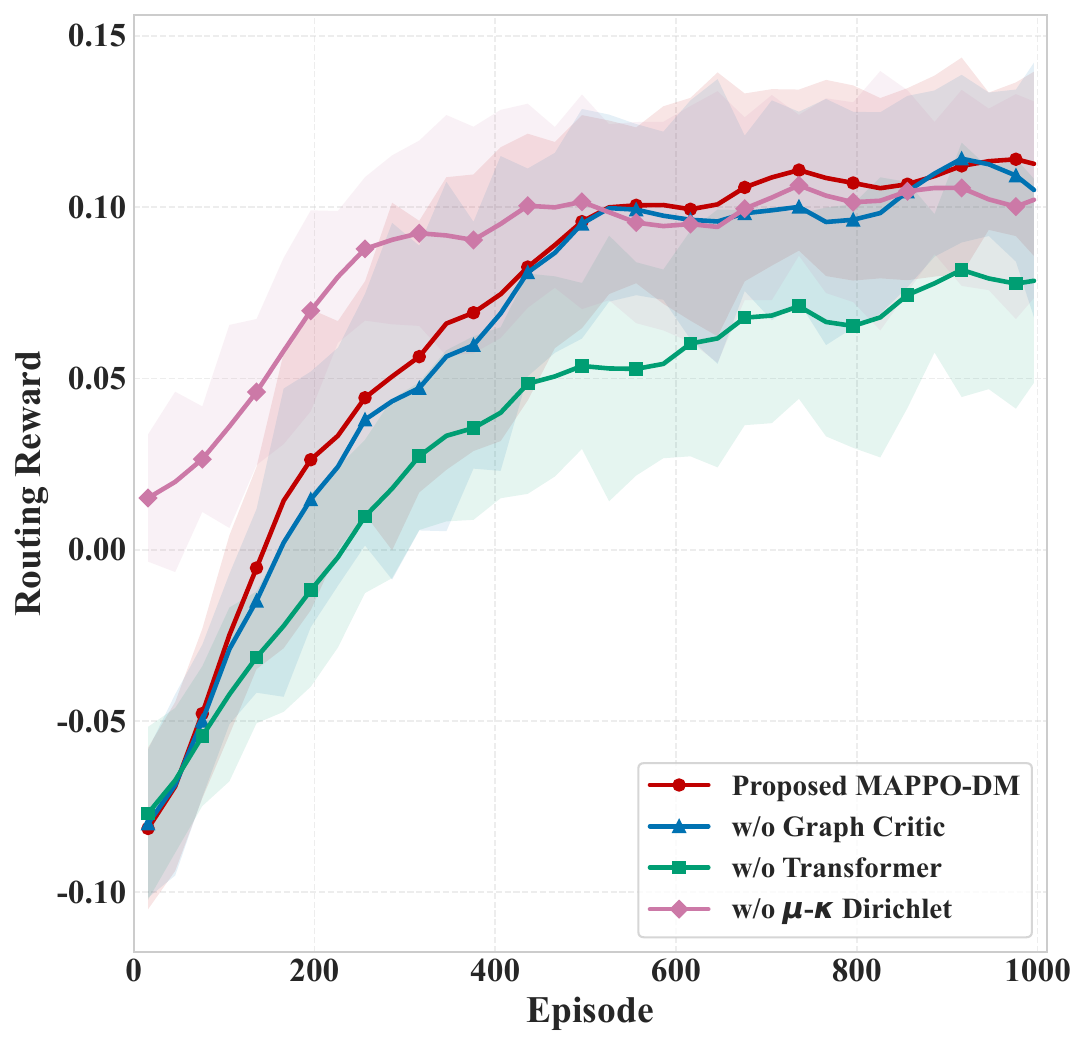}
    \caption{Convergence of the proposed algorithm and its ablated variants.}
    \label{convergence}
\end{figure}

\begin{figure}
    \centering
    \includegraphics[width=0.8\linewidth]{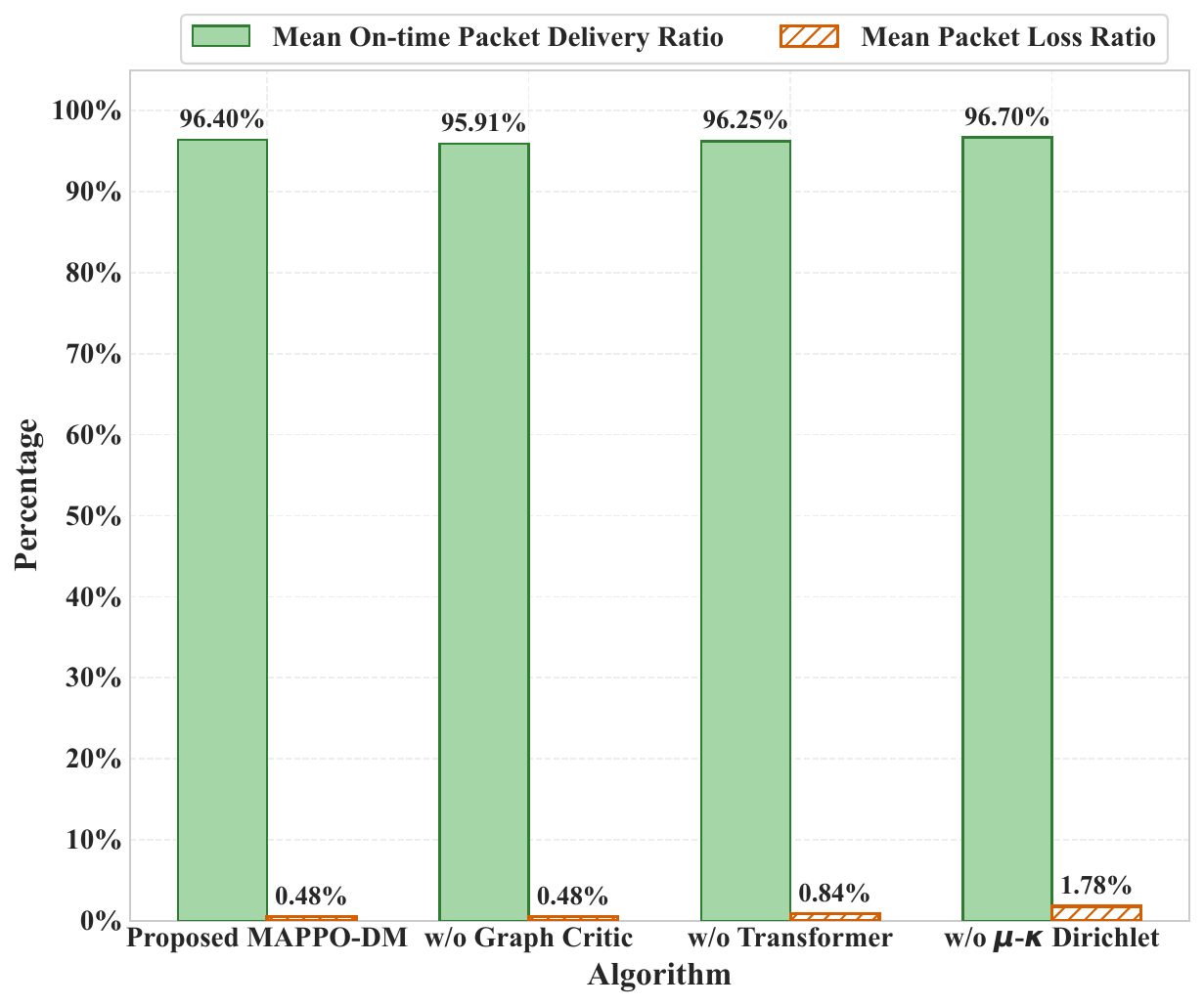}
    \caption{Performance of the proposed algorithm and its ablated variants.}
    \label{ablation_performance}
\end{figure}

Fig.~\ref{convergence} compares the training convergence of MAPPO-DM and its ablation variants, while Fig.~\ref{ablation_performance} compares their on-time packet delivery and packet loss ratios. The rewards of all methods gradually increase during training and eventually stabilize, indicating stable convergence. For on-time packet delivery, w/o $\mu$-$\kappa$ Dirichlet achieves the highest ratio of $96.70\%$, followed by MAPPO-DM, w/o Transformer, and w/o Graph Critic at $96.40\%$, $96.25\%$, and $95.91\%$, respectively. For packet loss, MAPPO-DM and w/o Graph Critic achieve the lowest ratio of $0.48\%$, compared with $0.84\%$ for w/o Transformer and $1.78\%$ for w/o $\mu$-$\kappa$ Dirichlet. These results show that the three components affect the two performance metrics differently. Removing the Graph Critic mainly reduces on-time delivery, whereas removing the Transformer also increases packet loss. Fixing $\kappa$ slightly improves on-time delivery but leads to a clear increase in packet loss, indicating the benefit of adaptive concentration control in reducing packet loss. Overall, the complete MAPPO-DM achieves a better balance between on-time packet delivery and packet loss.

\begin{figure}
    \centering
    \includegraphics[width=0.75\linewidth]{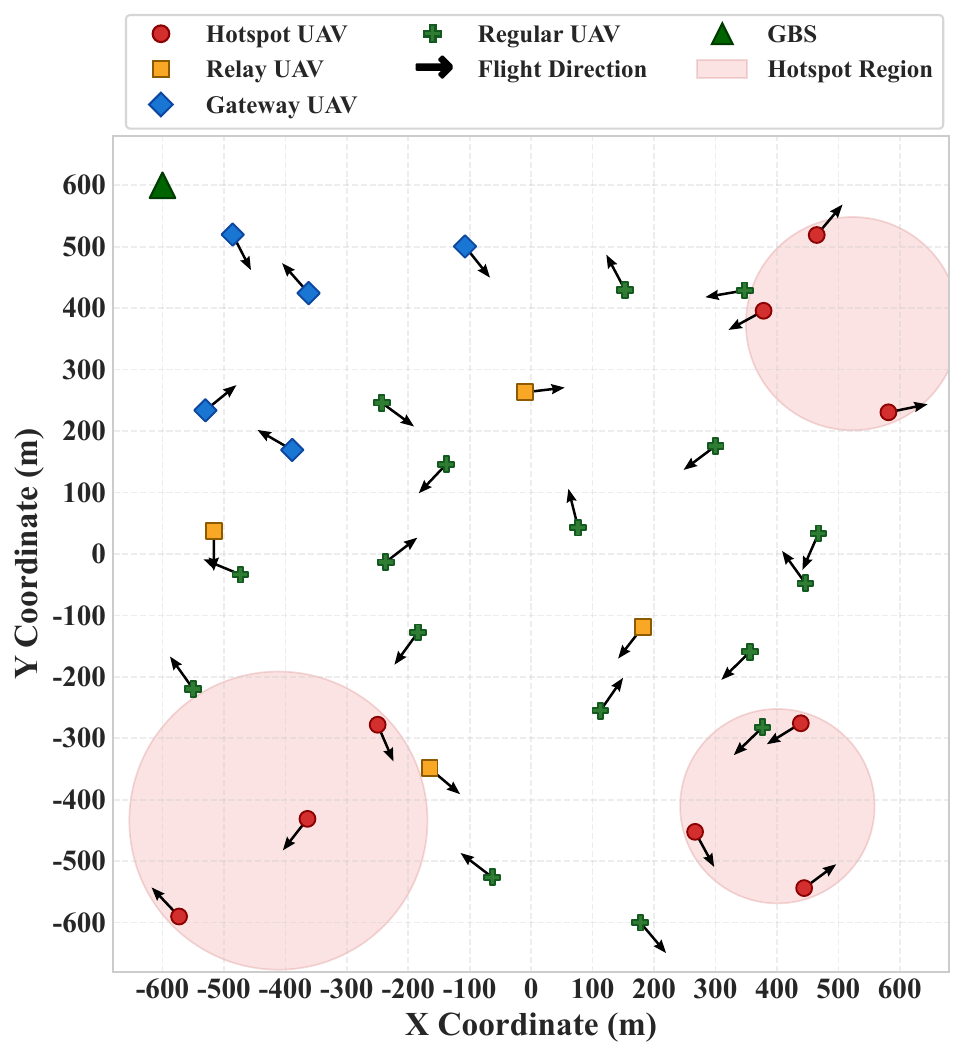}
    \caption{Spatial Distribution of GBS and UAV Trajectories.}
    \label{GBS-UAV-distri}
\end{figure}

Fig.~\ref{GBS-UAV-distri} shows the spatial layout of the simulation scenario, including the GBS, UAV locations, flight directions, and hotspot regions. The UAVs are divided into four roles, hotspot, relay, gateway, and regular UAVs. Regular UAVs make up the largest group, followed by hotspot UAVs, while relay and gateway UAVs are fewer. Regular and relay UAVs are distributed throughout the area, gateway UAVs are mainly located near the GBS, and hotspot UAVs are concentrated in the designated hotspot regions.\par

\begin{figure}
    \centering
    \includegraphics[width=0.8\linewidth]{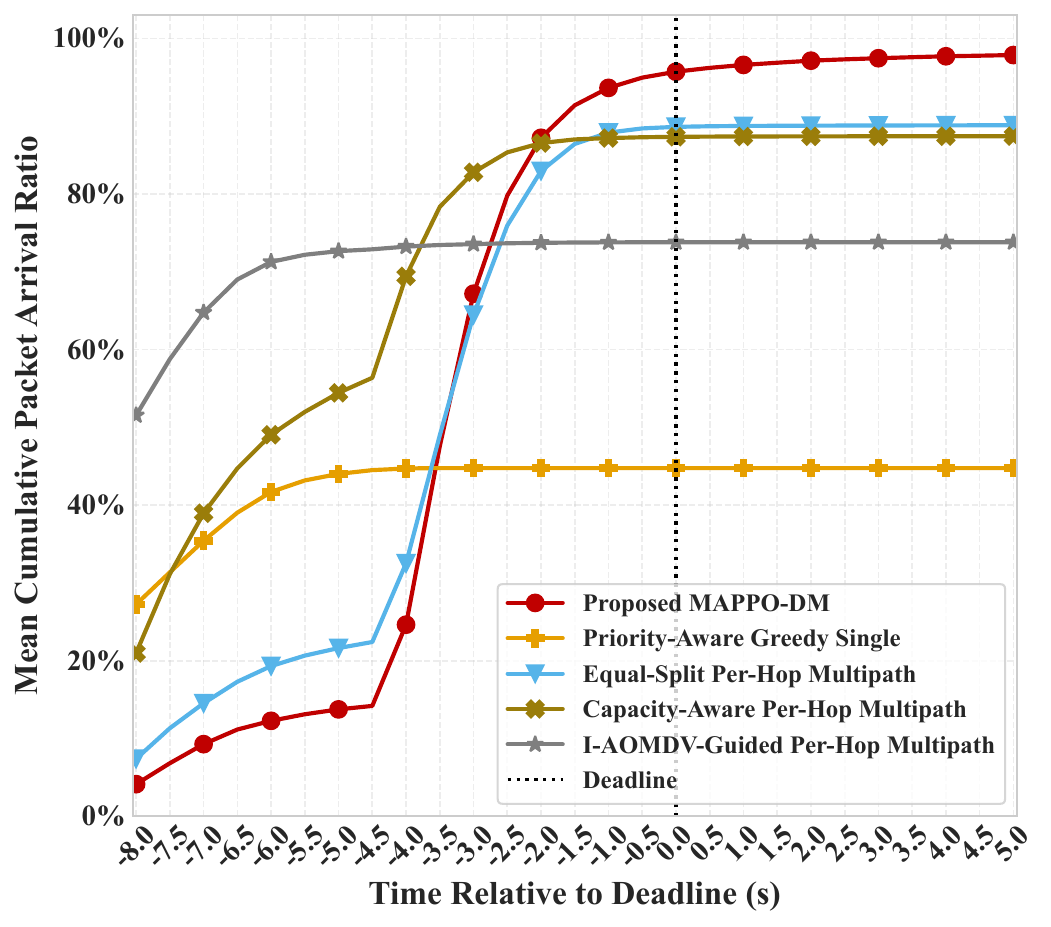}
    \caption{Cumulative packet arrival ratio.}
    \label{cumulative_arrival}
\end{figure}

Fig.~\ref{cumulative_arrival} shows the cumulative packet arrival ratios of different algorithms. The x-axis represents the arrival-time deviation, defined as the packet arrival time at the GBS minus its deadline. At each x-axis value, the y-axis gives the proportion of packets with deviations no greater than that value. For example, at $x=-1$, $0$, and $1$, the y-axis gives the cumulative fraction of packets that have arrived at least $1\ {\rm s}$ before their deadlines, by their deadlines, and within $1\ {\rm s}$ after their deadlines, respectively.\par

As shown in Fig.~\ref{cumulative_arrival}, Priority-Aware Greedy Single-Path cannot split traffic and is therefore more vulnerable to local congestion, resulting in the lowest cumulative arrival ratio of about $45\%$ . Among the multipath baselines, Capacity-Aware Per-Hop Multipath and I-AOMDV-Guided Per-Hop Multipath rise quickly for negative arrival-time deviations but level off near the deadline, reaching about $88\%$ and $75\%$ at $x=0$, respectively. Capacity-Aware Per-Hop Multipath mainly relies on local link capacity and neighbor queue states, without considering downstream conditions. I-AOMDV-Guided Per-Hop Multipath relies on periodically updated end-to-end path information and is less responsive to short-term network changes. Both methods are therefore more vulnerable to downstream congestion and packet loss. Equal-Split Per-Hop Multipath distributes traffic evenly among multiple next hops, which helps avoid traffic concentration. Its cumulative arrival ratio increases more gradually and reaches about $90\%$ at the deadline. In comparison, MAPPO-DM rises rapidly as the deadline approaches and surpasses all baselines before $x=0$. It reaches about $96\%$ at the deadline and nearly $98\%$ afterward. These results show that MAPPO-DM can adapt traffic allocation to traffic urgency and changing network conditions, allowing more packets to arrive before their deadlines. \par

\begin{figure}
    \centering
    \includegraphics[width=0.9\linewidth]{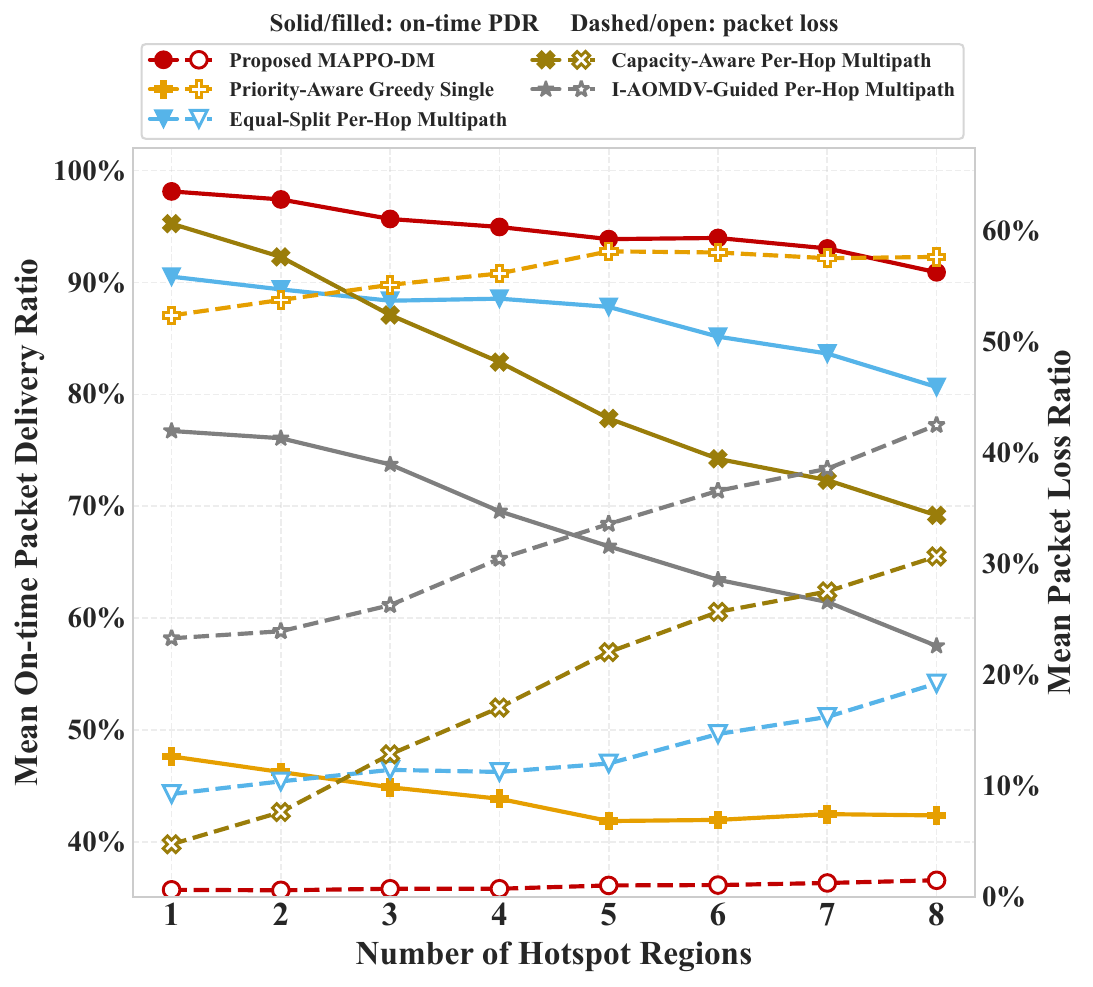}
    \caption{Performance under different numbers of hotspot regions.}
    \label{hotspot_performance}
\end{figure}

Fig.~\ref{hotspot_performance} compares MAPPO-DM with the baseline algorithms under different numbers of hotspot regions. The simulation area is divided into multiple equal-sized regions. For each setting, the specified number of regions is randomly selected as hotspot regions, excluding the region containing the GBS, and hotspot UAVs are deployed within these regions. As the number of hotspot regions increases, both the number of hotspot UAVs and the overall traffic load increase. Therefore, the on-time packet delivery ratio of all algorithms gradually decreases, while the packet loss ratio increases. Compared with the baseline algorithms, MAPPO-DM consistently achieves the best performance and shows a slower performance degradation as the traffic load increases. These results demonstrate that MAPPO-DM is more robust than other algorithms.

\begin{figure}
    \centering
    \includegraphics[width=0.85\linewidth]{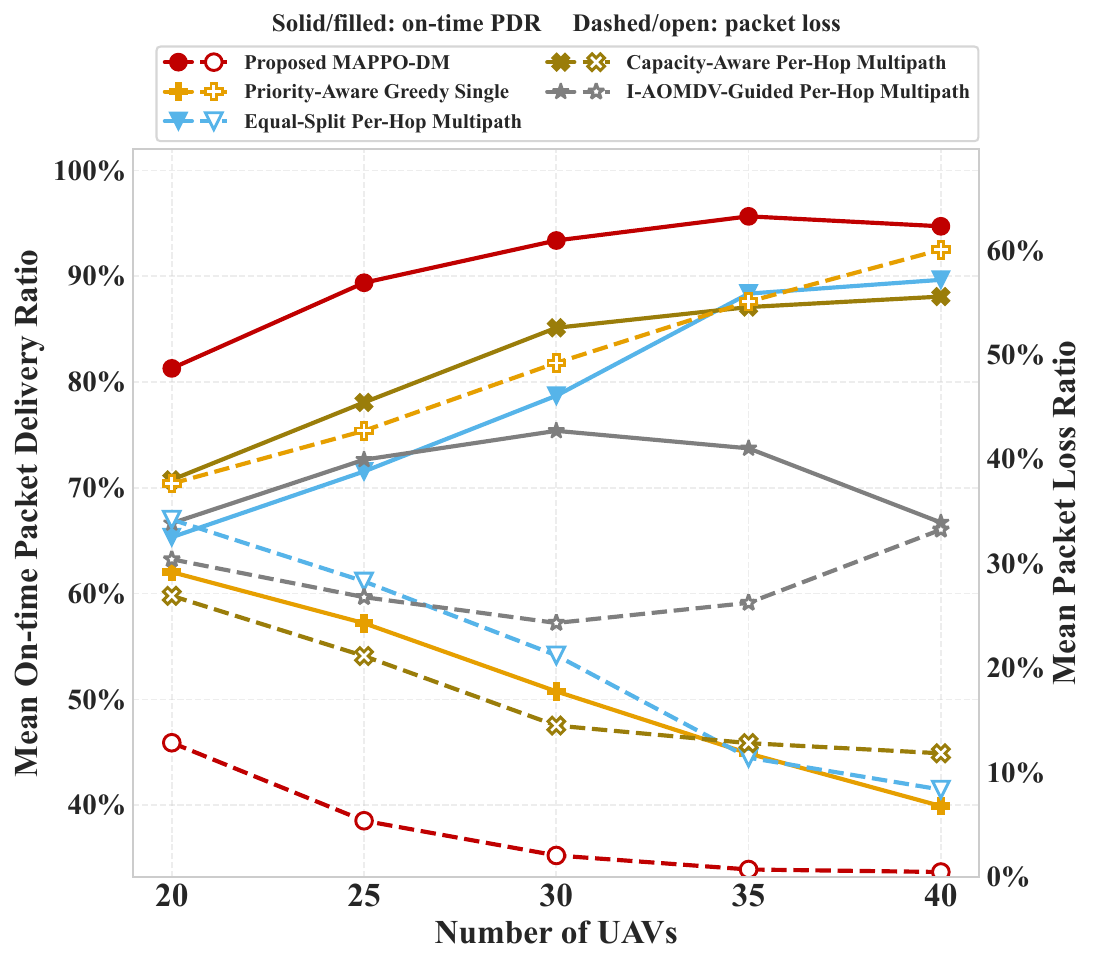}
    \caption{Performance under different numbers of UAVs.}
    \label{uavNumPDRDR}
\end{figure}

Fig.~\ref{uavNumPDRDR} compares the performance of different algorithms under different network sizes. To isolate the effect of network size, the overall traffic load is kept approximately constant. Specifically, as the number of UAVs increases, the task-generation probability of each UAV is reduced accordingly to maintain a similar aggregate task-generation rate.\par

As the network grows, the denser topology provides more forwarding options. MAPPO-DM, Equal-Split Per-Hop Multipath, and Capacity-Aware Per-Hop Multipath all benefit from these additional options, generally achieving higher on-time packet delivery ratios and lower packet loss ratios. When the number of UAVs increases from $35$ to $40$, their performance changes only slightly, indicating that further increasing network density provides limited benefits once sufficient forwarding options are available. I-AOMDV-Guided Per-Hop Multipath first improves and then degrades as the network size increases. At moderate network sizes, additional UAVs provide more end-to-end paths and improve routing flexibility. In denser networks, however, the larger number of links and paths makes end-to-end path information more difficult to maintain. Since I-AOMDV-Guided Per-Hop Multipath relies on periodic path updates, outdated information can reduce routing effectiveness. Priority-Aware Greedy Single-Path uses only one next hop and cannot balance traffic across multiple neighbors. As the topology becomes denser, multiple upstream UAVs may still select the same locally preferred next hops, causing traffic concentration and congestion. As a result, its on-time packet delivery ratio decreases and its packet loss ratio increases within the tested range.\par

\begin{figure}
    \centering
    \includegraphics[width=0.86\linewidth]{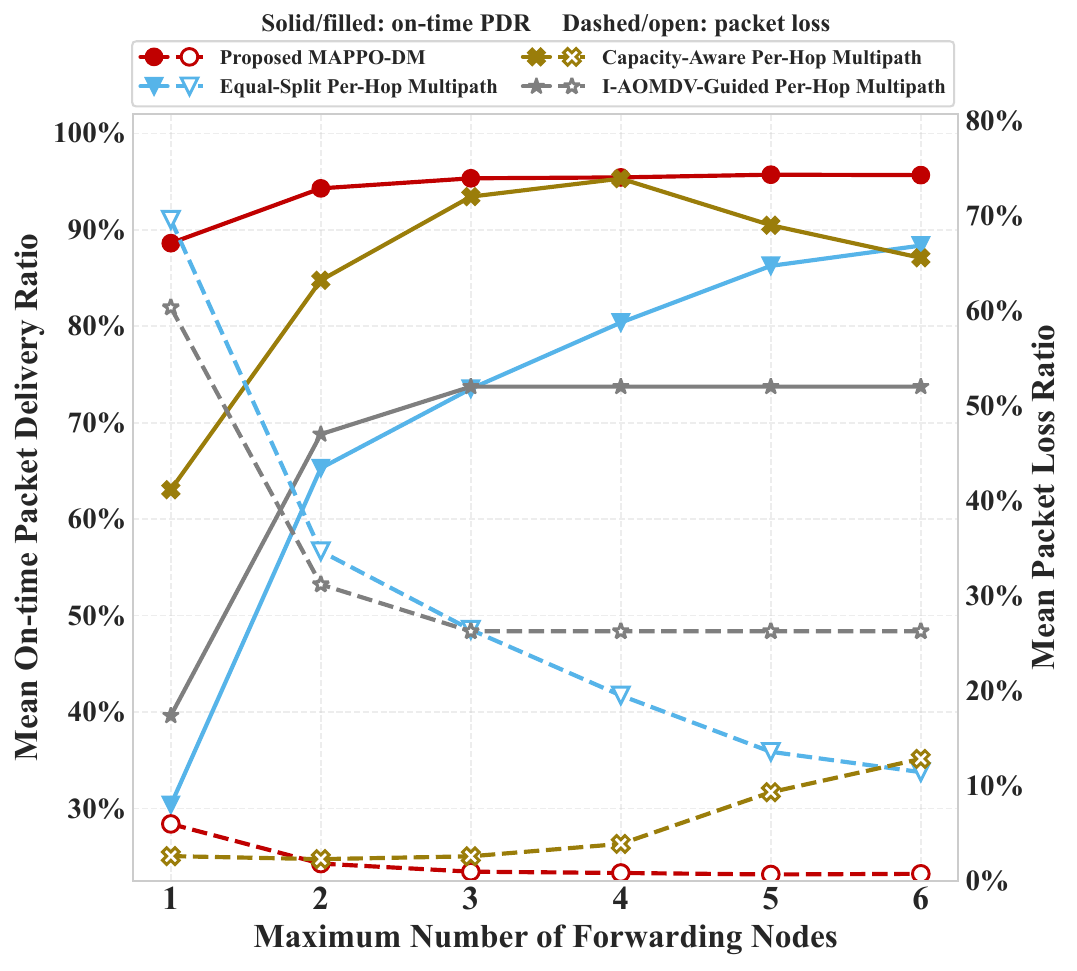}
    \caption{Performance under different numbers of candidate forwarding nodes.}
    \label{transNodePDRDR}
\end{figure}

Fig.~\ref{transNodePDRDR} shows the impact of the maximum number of forwarding nodes $N$ on the four multipath schemes. For Capacity-Aware Per-Hop Multipath, performance improves as $N$ increases from $1$ to $4$, since more high-capacity neighbors become available. However, when $N>4$, the on-time packet delivery ratio decreases and the packet loss ratio increases. This scheme mainly considers one-hop link capacity and receiver buffer availability. As $N$ increases further, more forwarding branches are activated and less traffic is retained locally. Although these additional neighbors can accept traffic at the current hop, their downstream forwarding conditions may be less favorable, leading to congestion and packet loss in later hops. I-AOMDV-Guided Per-Hop Multipath also benefits from a larger $N$ when the candidate set is small, but its performance becomes nearly stable once enough forwarding options are available. In comparison, MAPPO-DM is less sensitive to $N$. Its performance improves slightly from $N=1$ to $N=2$ and then remains nearly unchanged, showing that it can make effective routing decisions even with limited forwarding options. Equal-Split Per-Hop Multipath continues to improve as $N$ increases, since more forwarding branches allow traffic to be distributed more evenly and help reduce local congestion.\par

\section{Conclusions}
In this paper, a traffic-adaptive per-hop multipath routing framework was proposed, in which traffic can be dynamically split among multiple next-hop neighbors by each UAV. The routing problem was formulated to jointly improve the on-time packet delivery ratio and reduce the packet loss ratio, and the MAPPO-DM algorithm was developed to solve it. Simulation results show that the proposed algorithm achieves the best overall performance among the compared schemes. In particular, it maintains better latency guarantees and lower packet loss under different network loads, UAV scales, and forwarding constraints, demonstrating strong adaptability and robustness.\par

\section{Acknowledgment}
OpenAI ChatGPT was used for language editing and limited code assistance. All technical content and results were verified by the authors.


{
\bibliographystyle{IEEEtran}
\bibliography{IEEEabrv,bibe}
}

\end{document}